\documentclass[aps,prl,twocolumn,superscriptaddress]{revtex4-1}

\usepackage[english]{babel}
\usepackage{indentfirst}
\usepackage{graphicx}
\usepackage{subfigure}
\usepackage{algorithm}
\usepackage{algorithmicx}
\usepackage{algpseudocode}                        
\usepackage{amsmath}
\usepackage{caption}
\usepackage{enumerate}
\usepackage{appendix}
\usepackage[colorlinks,
            linkcolor=blue,
            anchorcolor=blue,
            citecolor=blue,
            urlcolor=blue
            ]{hyperref}

\begin{document}


\title{Multi-Matrix Post-Processing for Quantum Key Distribution}


\author{Chao-hui Gao}
\author{Dong Jiang}
\email{jiangd@nju.edu.cn}
\affiliation{State Key Laboratory for Novel Software Technology, Nanjing University, Nanjing, 210046, P.R.China}
\author{Liang-liang Lu}
\email{lianglianglu@nju.edu.cn}
\affiliation{School of Physics, Nanjing University, Nanjing, 210093, P. R. China}
\author{Yu Guo}
\author{Li-jun Chen}
\email{chenlj@nju.edu.cn}
\affiliation{State Key Laboratory for Novel Software Technology, Nanjing University, Nanjing, 210046, P.R.China}



\date{\today}

\begin{abstract}

Post-processing is a significant step in quantum key distribution(QKD), which is used for correcting the
quantum-channel noise errors and distilling identical corrected keys between two distant legitimate parties.
Efficient error reconciliation protocol, which can lead to an increase in the secure key generation rate, is one of the main performance indicators of QKD setups. In this paper, we propose a multi-low-density parity-check
codes based reconciliation scheme, which can provide remarkable perspectives for highly efficient
information reconciliation. With testing our approach through data simulation, we show that the proposed scheme combining multi-syndrome-based error rate estimation allows a more accurate estimation about the error rate as compared with random sampling and single-syndrome estimation techniques before the error correction, as well as a significant increase in the efficiency of the procedure without compromising security and sacrificing
reconciliation efficiency.
\end{abstract}


\maketitle


\section{Introduction}

Quantum Key Distribution (QKD) is a class of protocols where the two separated users, Alice and Bob, can share identical secret keys which are secure from the eavesdropper (Eve) \cite{gisin2002quantum}.
Since it provides unconditional security guaranteed by laws of quantum mechanics \cite{scarani2009security}, QKD has attracted wide attention and many advanced works have been published over recent years \cite{lo2005decoy,lo2012measurement,liao2017satellite,Wang2005Beating}.
Generally, a QKD protocol can be divided into quantum and classical parts.
In the former part, Alice generates and transmits a set of raw key through the quantum channel.
Due to Eve's attacks\cite{Bennet1984Quantum}, channel noise, and device imperfection \cite{Gerhardt2010Full, Weier2011Quantum, Jain2011Device}, the keys are weakly correlated and partially secure, and Eve may obtain some information about the keys.
The classical part, also known as post-processing, is used to correct the errors, and to remove information leakage.

Post-processing consists of base sifting \cite{Bennet1984Quantum}, error estimation \cite{Wang2005Beating,treeviriyanupab2014rate,2018arXiv181005841K}, key reconciliation \cite{Luby1998Improved} and privacy amplification \cite{Bennett1988Privacy,Bennett1995Generalized}.
During base sifting, the bits measured with correct measurement bases in the raw key are kept and constitute the sifted key.
Subsequently, Bob uses a key reconciliation algorithm to correct the errors in the sifted key based on the estimated error rate.
Finally, Alice and Bob implement privacy amplification to remove information leakage and obtain the final key, which is secure from Eve.

In error estimation, the accuracy of the estimated quantum bit error rate(QBER) effects the operational efficiency of post-processing.
If the actual QBER for a given block is larger than the estimate, Bob might end up with a wrong final key.
A common method to obtain the QBER for legitimate users is to exchange and compare random sampled sifted key, which can lower the
key generation rate due to disclosed bits. Recently, Kiktenko $et al$ \cite{2018arXiv181005841K} proposed a distinct approach
based on the use of syndromes of low-density parity-check (LDPC) codes to obtain the QBER for each block of the sifted key,
allowing more accurate estimation. The suggested algorithm is also suitable for irregular LDPC codes.

In parallel, key reconciliation is the most crucial step of post-processing, which is responsible for correcting the errors in Bob's sifted key,
in such a way that it ensures consistency between Alice's and Bob's sifted keys.
Belief Propagation (BP) \cite{Luby1998Improved} is the most widely used key reconciliation algorithm, and has attracted intensive study \cite{Kou2001Low,Zhang2002Shuffled,Hocevar2004,Sharon2004An,Zhang2004A,Chang2008Lower,Park2009Shuffled,Wu2010Alternate,Aslam2017Edge}.
There are three criteria for judging a key reconciliation algorithm, namely, convergence speed, bit error rate (BER) and success rate.
However, it is hard to meet the three criteria at the same time, which often appears if the syndrome decoding, based on
an iterative BP algorithm, fails to converge
within the predefined number of iterations (e.g., it could
be caused by an inappropriate choice of the LDPC parity-matrices
relative to the actual errors in raw keys).
This makes key reconciliation the bottleneck of QKD and severely affects the key generation rate for industrial QKD systems.

In this paper, we extend the blind information reconciliation \cite{kiktenko2017symmetric} to multiple LDPC codes and estimate the QBER more
accurately by virtue of multiple syndromes without disclosing redundant bits.
Experimental results show that a significant increase in the efficiency of the procedure, i.e. faster convergence speed with higher success rate.
To prevent extra information leakage in our post-processing scheme, we also give a multiple LDPC codes construction method.
Security analysis shows that our key reconciliation scheme does not reveal extra information.

The rest of the paper is organized as follows: in Section \uppercase\expandafter{\romannumeral2}, a briefly review of error estimation and key reconciliation is given, followed by a detail description of the process and advantages of our scheme. Section \uppercase\expandafter{\romannumeral3}
provides the novel multi-matrix post-processing approach for error estimation and correction. In Section \uppercase\expandafter{\romannumeral4} a set of data simulation are carried out to fully evaluate these advantages. The proposed construction method of multiple matrices and the security analysis of the proposed scheme are given in the appendix.

\section{Preliminaries}
\label{sec:examples}

In this section, we will first review error estimation and reconciliation. Other parts of post-processing can be referred to \cite{Bennet1984Quantum,Bennett1988Privacy,Bennett1995Generalized}.

%
%

\subsection{Error Estimation}

We assume that Alice and Bob possess random sifted keys of equal length, and Bob needs to estimate the error rate $e$ of the sifted keys before executing key reconciliation, since $e$ is an important input parameter of reconciliation algorithms. The estimation accuracy of $e$ directly effects the operational efficiency of post-processing.
If $e$ is overestimated, Alice will place superfluous information on her syndrome, i.e., more leakage needed to be removed during privacy amplification, leading to relatively low key generation rate.
On the contrary, if $e$ is underestimated, less information is provided, so Bob spends more time to correct errors during key reconciliation or even end up with wrong final key.

\begin{figure}[htbp]
    \raggedleft
    \includegraphics[width=0.46\textwidth]{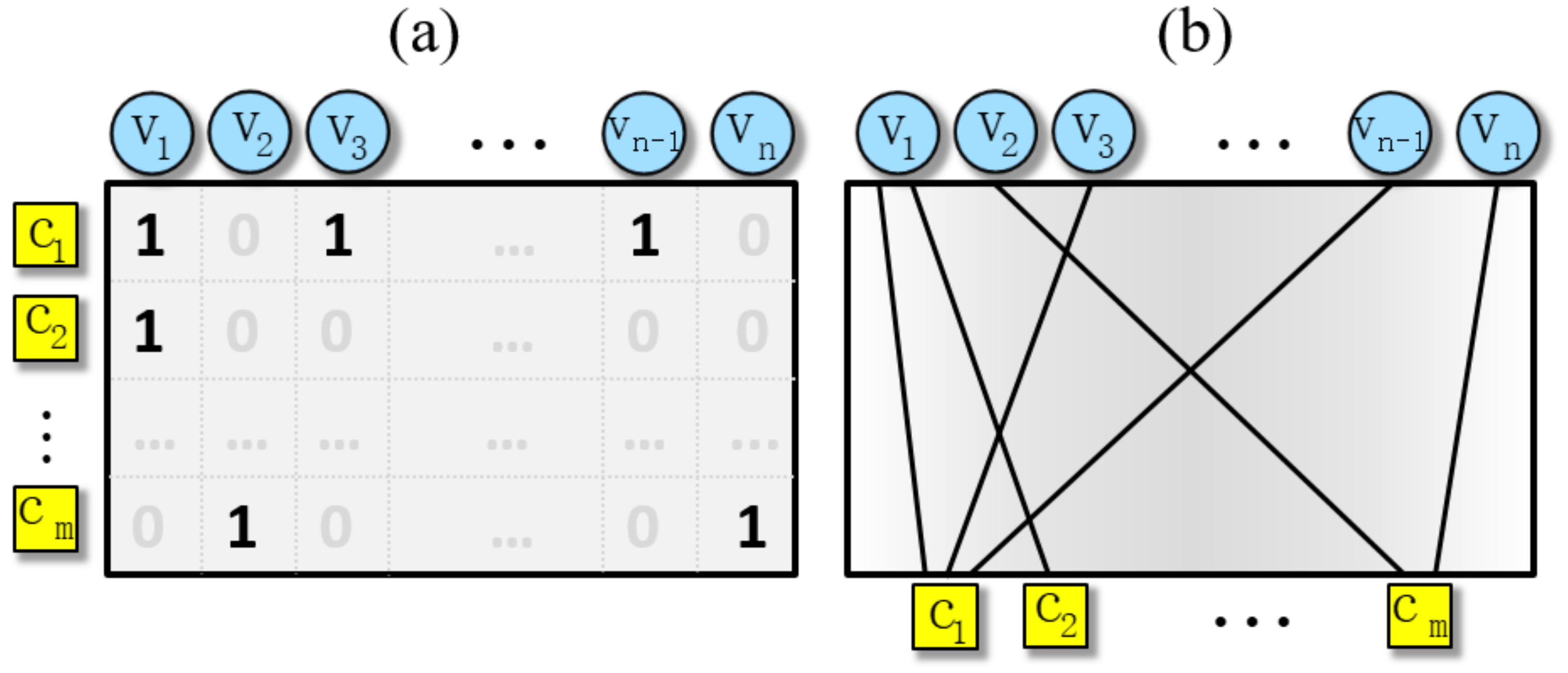}
    \caption{A binary m $\times$ n LDPC matrix (a) and its corresponding TG (b).}
	\label{fig:LDPC-TG}
\end{figure}

Error estimation can be executed in the several ways.
The most well-known method is the random sampling \cite{Wang2005Beating}.
But its drawback is that if Alice and Bob want to estimate more accurate error rate, they inevitably sacrifice key bits.
To solve this problem, P.Treeviriyanupab \emph{et al}. proposed a new method \cite{treeviriyanupab2014rate}.
In this protocol, Alice and Bob use their syndromes $z^{A}=[z_1^{A}, \cdots, z_m^{A}]$ and $z^{B}=[z_1^{B}, \cdots, z_m^{B}]\ (z_j^{A},z_j^{B}\in \{0,1\}, j\in\{0,\cdots,m\})$ as input to calculate the maximum likelihood estimation of error rate.
Syndromes are generated from a kind of data structure, LDPC code \cite{Gallager1962Low}, which can be presented by a $m\times n$ matrix or a Tanner Graph (TG) \cite{Tanner1981A}.
In Fig. \ref{fig:LDPC-TG} (a), an example of binary LDPC matrix $H_{m\times n}$ is given.
The variable nodes $v_i~(i\in\{1,\cdots,n\})$ (blue circles ) and check nodes $c_j~(j\in\{1,\cdots,m\})$ (yellow squares) represent bits of key and parity-check equations, respectively \cite{Gallager1962Low}.
TG corresponding to this matrix is shown in Fig. \ref{fig:LDPC-TG} (b).
An edge connecting a variable node and a check node indicates that the variable node participates in the parity-check equation.
In a LDPC code, the degree of a variable node (or check node) is the number of check nodes (or variable nodes) connected to it.
The syndromes, $z^{A}$ (or $z^{B}$), are simply obtained by multiplying a LDPC matrix and Alice's (or Bob's) sifted key.
But the method \cite{treeviriyanupab2014rate} is applicable only to regular LDPC code, in which all of the variable nodes have the same degrees and so does all check nodes. So Kiktenko \emph{et al}. extend the scope of application \cite{2018arXiv181005841K}  (hereinafter referred to as the single-syndrome error estimation), which is also suitable for irregular LDPC code.

\subsection{Key Reconciliation}
BP \cite{Luby1998Improved}, also known as the Sum Product (SP) algorithm, can be used for error-correction.
Due to its relatively high decoding efficiency and low executing complexity, BP has been widely adopted in QKD to correct the key errors caused by Eve's attacks, channel noise, etc.

In QKD, if Bob uses BP to correct his sifted key $y^{T}=[y_1,\ldots,y_n]$, he first needs to initializes $P^{b}_i~(b\in\{0,1\})$, $v_i$  and variable-to-check (V2C) information $L_{v_i\to c_j}$ as follows,
\begin{equation}
\begin{cases}
P^{0}_i=1-e, P^{1}_i=e & y_i=0\\
P^{0}_i=e, P^{1}_i=1-e & y_i=1
\end{cases},
\label{equ:bp_initial}
\end{equation}

\begin{equation}
L_{P_i}=\log \frac{P^{0}_i}{P^{1}_i},
\label{equ:L-Pi}
\end{equation}

\begin{equation}
L_{v_i\to c_j}=L_{P_i},
\label{equ:init-Lq}
\end{equation}
\noindent where $P^{b}_i~(b\in\{0,1\})$ is the prior probability of the candidate value $b$ of $v_i$, $e$ is the result of error estimation, $L_{P_i}$ represents the log likelihood ratio of $P_i^b$.

\begin{figure}[htbp]
	\centering
    \includegraphics[width=0.46\textwidth]{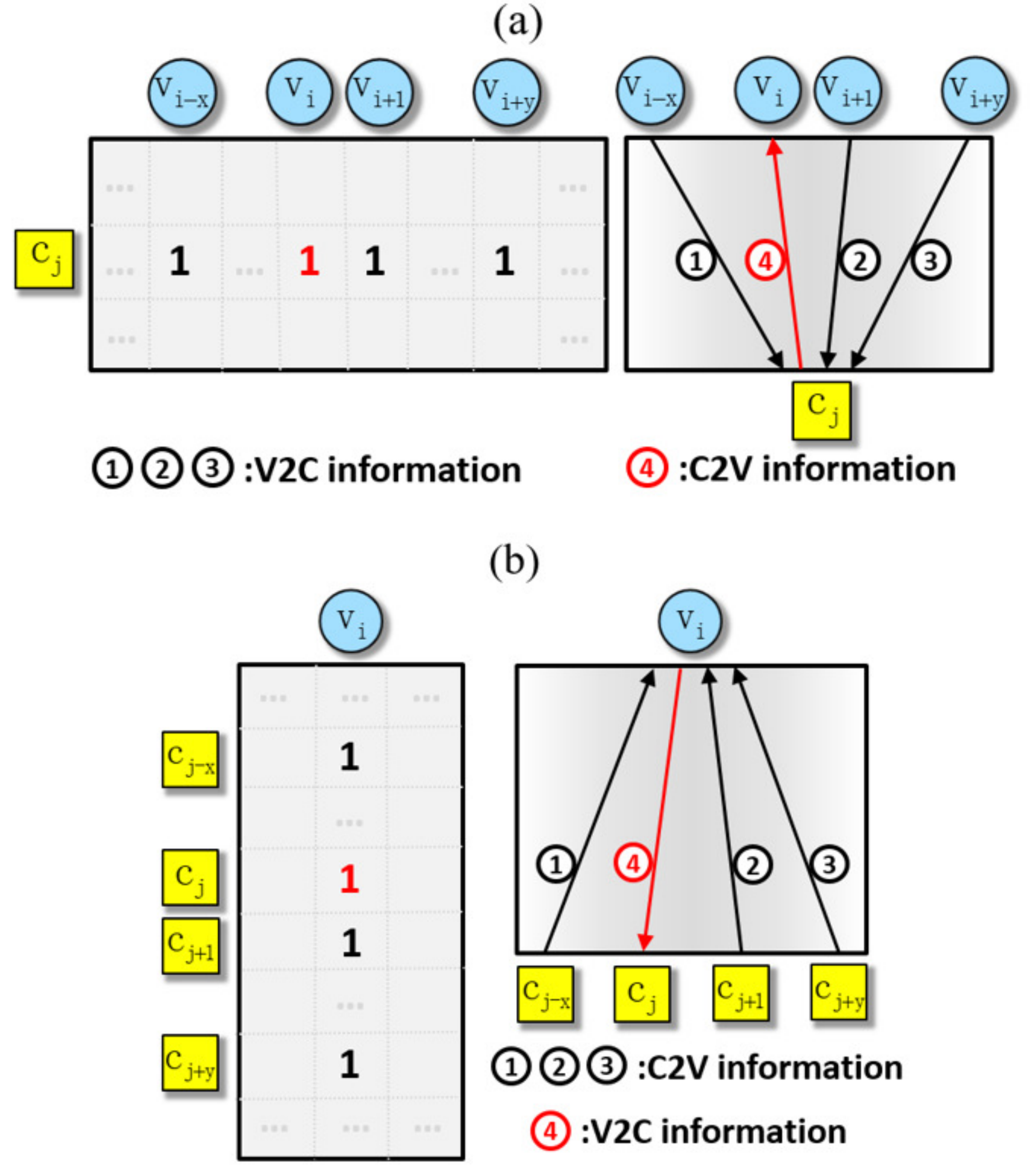}
	\caption{(a)Generated C2V information $L_{c_j\to v_i}$ by Alice. (b) Updated V2C information $L_{v_i\to c_j}$ by Bob.}
	\label{fig:C2VandV2C}
\end{figure}

Secondly, as shown in Fig.  \ref{fig:C2VandV2C} (a), he generates and propagates check-to-variable (C2V) information $L_{c_j\to v_i}$ by
\begin{equation}
\begin{split}
L_{c_j\to v_i}=\mathrm{sign}(z_j)\cdot 2\mathrm{tanh}^{-1}(\prod_{v_{i^{'}}\in{N(c_j)\backslash i}}\mathrm{tanh}(\frac{1}{2}L_{v_{i^{'}}\to c_j})),
\label{equ:L-rji}
\end{split}
\end{equation}
\noindent where $z$ denotes the Alice's syndrome \cite{Mackay1999Good}, which is the product of $H_{m\times n}$ and Alice's sifted key, $\mathrm{tanh()}$ is the hyperbolic tangent function, $\mathrm{tanh}^{-1}()$ is the inverse function of $\mathrm{tanh}()$, $v_{i^{'}}\in{N(c_j)\backslash i}$ represents the set of adjacent variable nodes of check nodes $c_j$ except $v_i$, $\mathrm{sign}()$ is a sign function defined as follows:
\begin{equation}
\mathrm{sign}(z_j)=
\begin{cases}
+1& z_j=0 \\
-1& z_j=1
\end{cases}.
\end{equation}

Thirdly, as plotted in Fig. \ref{fig:C2VandV2C} (b), Bob updates and propagates V2C information by substituting the generated C2V information into the following equation.
\begin{equation}
L_{v_i\to c_j}=L_{P_i}+\sum_{c_{j^{'}}\in{N(v_i)\backslash j}}L_{c_{j^{'}}\to v_i},
\label{equ:L-qij}
\end{equation}
where, $c_{j^{'}}\in{N(v_i)\backslash j}$ represents the set of adjacent check nodes of $v_i$ except $c_j$. All of $L_{c_j\to v_i}$ and $L_{v_i\to c_j}$ contain information of posterior probabilities of $v_i$.

Finally, he calculates the soft-decision value of every variable node $v_i$ as follows,
\begin{equation}
L_{v_i}=L_{P_i}+\sum_{c_j\in{N(v_i)}}L_{c_j\to v_i},
\end{equation}
then performs the decoding decision on every variable node according to the following equation,
\begin{equation}
y_i=
\begin{cases}
1& L_{v_i}>0 \\
0& L_{v_i}<0
\end{cases}.
\label{equ:decode-decide}
\end{equation}
Bob iterates the last three steps until the decoding is successful (i.e., the equation $z=H_{m\times n }\cdot y$ is satisfied) or the number of iterations reaches the pre-set upper limit.

In each iteration, BP can use different scheduling strategies, which can be divided into three categories \cite{Casado2007Informed}: Flooding, Shuffled, and Layer.
Flooding first goes through all the check nodes and generates C2V information, then traverses all the variable nodes and updates V2C information. Shuffled uses variable nodes as the traversal sequence, sequentially updates C2V and V2C information between variable nodes and their adjacent check nodes.
Layer, on the contrary, uses check nodes as the traversal sequence, sequentially updates C2V and V2C information between check nodes and their adjacent variable nodes.
In practical applications, BP, Shuffled Belief Propagation (SBP) \cite{Zhang2002Shuffled}, and Layer Belief Propagation (LBP) \cite{Hocevar2004,Sharon2004An} are the typical representatives of the above three scheduling strategies. For convenience, the algorithms based on single matrix are hereinafter referred to as the single-matrix reconciliation.

%

\section{Multi-matrix Post-processing}

In this section, we propose a post-processing scheme where users estimate error rate with multiple syndromes and correct errors with multiple matrices (hereinafter referred to as the multi-matrix post-processing). In the multi-matrix post-processing, base sifting and privacy amplification are the same as the original post-processing (hereinafter referred to as the single-matrix post-processing). Here we introduce only error estimation and key reconciliation in the frame of multiple syndromes.

\subsection{Multi-syndrome Error Estimation}

Each bit of a syndrome represents the relationship of the parity-check equation and the key. By comparing Alice's syndrome and his own syndrome, Bob can extract some information about error rate. If he uses multiple matrices, he can obtain multiple syndromes, which can be used to estimate the error rate more accurately.

Above all, Bob obtains $u$ syndromes from Alice and performs XOR as follows,
\begin{equation}
\triangle z^{k}=z^{A|k} \oplus z^{B|k},k\in \{1,\cdots,u\},
\label{equ:syndromes}
\end{equation}
where $\oplus$ is the XOR operation, $z^{A|k}$ and $z^{B|k}$ is the $k^{th}$ syndromes of Alice and Bob respectively. Then Bob calculates the maximum likelihood estimation of $e$ by,
\begin{equation}
e=\arg \max_{e^{'}\in[0,threshold]}M_{e^{'}|\triangle Z},
\label{equ:errorRates}
\end{equation}
where $e^{'}$ is a possible value that $e$ may take, $\triangle Z=[\triangle z^{1},\triangle z^{2},\cdots ,\triangle z^{u}]$. In equation (\ref{equ:errorRates}), $M_{e^{'}|\triangle Z}$ can be obtained via,
\begin{equation}
M_{e^{'}|\triangle Z}=\prod_{k=1}^{u}\prod_{j=1}^{m}[1-\triangle z_j^{k}+(2\triangle z_j^{k}-1)p(e^{'},d_{c_j}^{k})],
\end{equation}

\begin{equation}
\begin{split}
p(e^{'},d_{c_j}^{k})&=Pr(\triangle z_j^{k}=1)\\&
=\sum_{\substack{i=1\\ i\bmod 2=1}}^{d_{c_j}^{k}}\dbinom{d_{c_j}^{k}}{i}e^{'i}(1-e^{'})^{d_{c_j}^{k}-i},
\label{equ:probabilities}
\end{split}
\end{equation}
where $M_{e^{'}|\triangle Z}$ is the likelihood function of $e^{'}$, $p(e^{'},d_{c_j}^{k})$ is the priori probability of that $z_j^{A|k}$ and $z_j^{B|k}$ are different, $\triangle z_j^{k}$ is the $j^{th}$ bit of $\triangle z^{k}$, $z_j^{A|k}$ is the $j^{th}$ bit of $z^{A|k}$, $z_j^{B|k}$ is the $j^{th}$ bit of $z^{B|k}$, $d_{c_j}^{k}$ is the degree of check node $c_j$ of $k^{th}$ matrix. As shown in equation (\ref{equ:errorRates}), $e$ evaluates to $e^{'}$ that maximizes $M_{e^{'}|\triangle Z}$. The \lq\lq threshold\rq\rq\ \cite{richardson2001capacity,richardson2001design} is the upper limit of error rate that can be acceptable. If $e$ exceeds the \lq\lq threshold\rq\rq, the sifted key will be abandoned.

Our method (hereinafter referred to as the multi-syndrome error estimation) is based on the single-syndrome error estimation, but can bring out higher accuracy of estimation. Meanwhile, compared with the random sampling, our method doesn't need to discard any key bit.

%
%
%

\subsection{Multi-matrix Key Reconciliation}

Although, theoretical analysis and simulation results show that the single-matrix reconciliation can correct the errors to some extent \cite{Sharon2007Efficient},
the performances of convergence speed and BER are still limited \cite{Casado2007Informed,Casado2010LDPC}, and the success rate is decreased when LDPC code is not cycle-free \cite{Tanner1981A,Yazdani2004Improving}.
To overcome these problems, we propose a new reconciliation strategy that uses two or more matrices to correct errors in parallel.
Let us take multi-matrix BP (MBP) as an example to show the detailed process and advantages of our strategy.

Suppose Alice and Bob have prepared and shared $u$ LDPC codes $H_1, \ldots, H_u$. After obtaining the sifted key $x^{T}=[x_1, \ldots, x_n]~(x_i\in\{0,1\})$, Alice calculates $u$ syndromes according to the following equation:
\begin{equation}
\begin{split}
(z^k)^T=[z_1^k,\ldots,z_n^k]=H_k\cdot x,~ &k\in\{1,\dots,u\},\\
																			   &z_i^k\in\{0,1\},
\end{split}
\end{equation}
\noindent and sends them to Bob over the classical channel. Because of Eve's attacks, channel noise, or device imperfection,
Bob inevitably obtain different sifted keys with Alice, denoted as $y^{T}=[y_1, \ldots,y_n],~(y_i\in\{1,0\})$.

In our strategy, Bob first initializes the prior probabilities $P^{b}_i~(b\in\{0,1\})$, log likelihood ratios $L^k_{P_i}$ and V2C information $L^k_{v_i\to c_j}$ for all matrices according to equations  (\ref{equ:bp_initial}), (\ref{equ:L-Pi}) and (\ref{equ:init-Lq}), respectively.

Secondly, Bob generates and propagates C2V information $L^{k}_{c_j\to v_i}$ according to equation (\ref{equ:L-rji}).

Thirdly, by substituting C2V information into equation (\ref{equ:L-qij}), Bob updates and propagates V2C information.

Finally, he goes through all variable nodes to obtain their soft-decision values by
\begin{equation}
L_{v_i}=L_{P_i}+\sum^{u}_{k=1}\sum_{c_j\in{N_{k}(v_i)}}{L^k_{c_j\to v_i}},
\label{equ:M-decision}
\end{equation}
\noindent and makes decoding decisions according to equation (\ref{equ:decode-decide}).
Because once Bob's key is corrected, i.e. $y$ is equal to $x$, all his syndromes satisfy $z_k=H_k\cdot y$. Thus he randomly selects a matrix $H_k$, and judges whether $z_k$ is equal to $H_k\cdot y$. If so, Bob terminates the algorithm and stores $y$. Otherwise, he starts another iteration. The reconciliation is considered as a failure when the number of iterations exceeds the upper limit.

There is an important figure called the reconciliation efficiency $f$ \cite{kiktenko2017symmetric}. It shows the ratio of practical information leakage to theoretical floor for successful reconciliation. It serves to imply the efficiency and security of a reconciliation strategy and help privacy amplification to remove information leakage. For the single-matrix reconciliation, the reconciliation efficiency $f$ is represented as
\begin{equation}
f=\frac{m}{nh(e)}>1,
\label{equ:f-single}
\end{equation}
where m and n are the numbers of check nodes and variable nodes of the LDPC code, e is the result of error estimation, h is the Shannon binary entropy:
\begin{equation}
h(e)=-e\log_2e-(1-e)\log_2(1-e).
\label{equ:entropy}
\end{equation}
For the multi-matrix reconciliation, however, $f$ is given by
\begin{equation}
f=\frac{\alpha m}{nh(e)}>1,(\alpha\geq 1),
\label{equ:f-multi}
\end{equation}
where $\alpha$ is a constant which is relative to $u$ and the structures of $u$ matrices. Fortunately, if the construction method of multiple matrices (see Appendix B) is used, it can be proved that the practical information leakage is equal to $m$ (see Appendix A), i.e., $\alpha$ is equal to $1$, without sacrificing the reconciliation efficiency compared
with single-matrix post-processing.


Obviously, our strategies is portable, it can be easily applied to SBP, LBP (see Appendix C), and other algorithms to achieve the following improvements:

\begin{enumerate}[1.\itemindent=-\itemindent]
\item\textbf{ Faster Convergence Speed} \ \ In our strategy, when Bob generates C2V and updates V2C information, all matrices operate in parallel. And as shown in equation (\ref{equ:M-decision}), Bob obtains the soft-decision value of each variable node $v_i$ by gathering all the C2V information sent to $v_i$ in every matrix. The amount of C2V information gathered within one iteration in the multi-matrix reconciliation is equal to information gathered in numerous iterations in the single-matrix reconciliation.	
\item\textbf{ Higher Success Rate} \ \ Once C2V and V2C information of a matrix are trapped in a cycle, the other matrices without this cycle can help the trapped matrix jump out the cycle, leading to higher success rate.
\item\textbf{ Lower BER} \ \ The value of each key bit is determined according to the information provided by multiple matrices. The accuracy of error-correction is effectively improved, resulting in lower BER.
\end{enumerate}

\section{Experimental Evaluation}
To fully evaluate the above advantages of multi-matrix post-processing, in this session we first give some detailed comparisons among three methods of error estimation.
Then for the other three parts, the experiments about the three criteria of key reconciliation algorithms are carried out.
All simulation data used in our experiments are generated by real random number generator IDQ EasyQuantis 2.1.
For comparison, we also set the upper limit of iterations to $100$, which is similar to existing implementations \cite{Zhang2012Verification,Djordjevic2012Evaluation}, and the code rate and code length of LDPC codes are set to $0.8$ and $10000$, respectively.

\subsection{Error Estimation}

\begin{figure*}[htbp]
	\centering
    \includegraphics[width=1\textwidth]{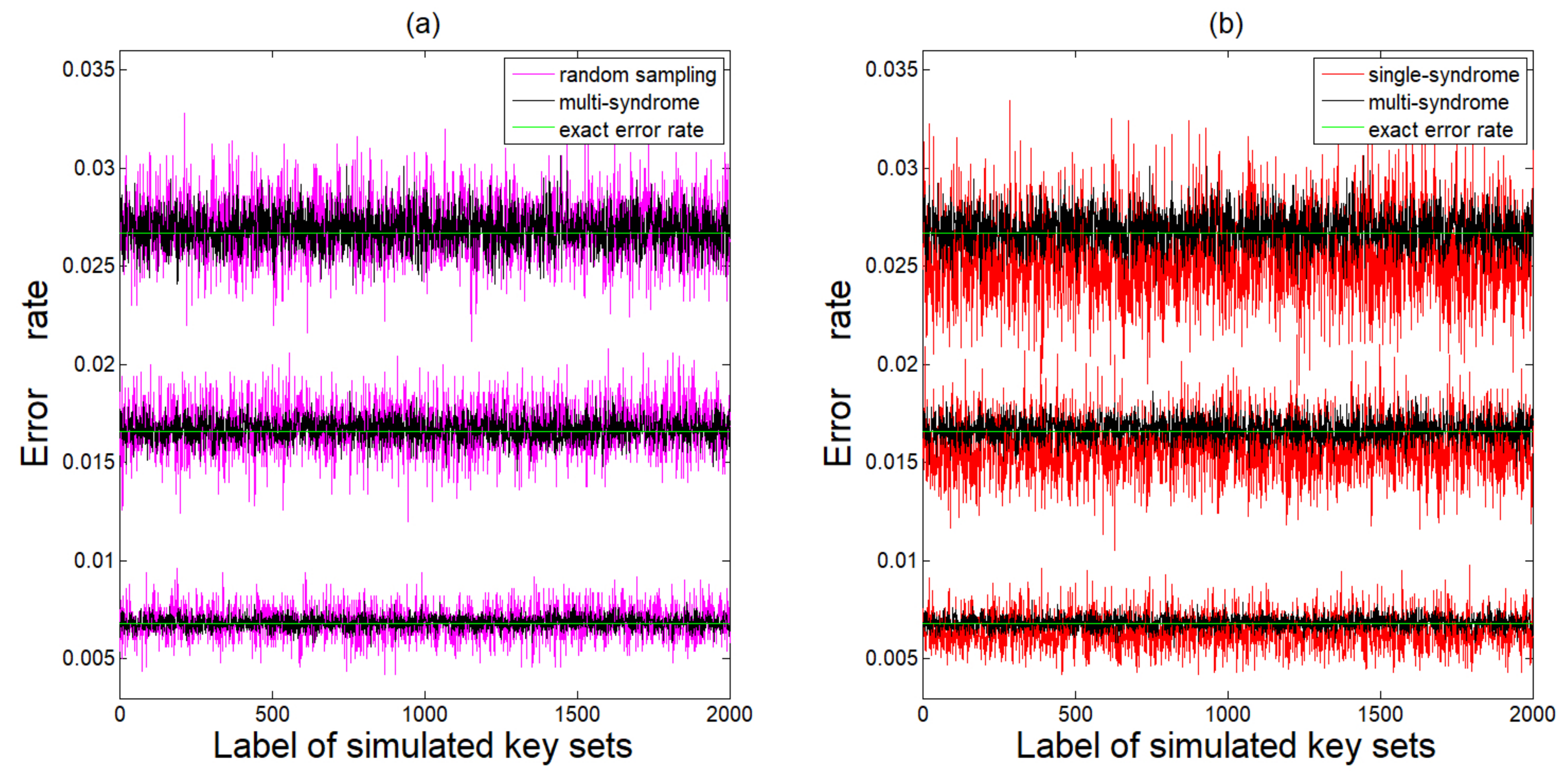}
	\caption{ Comparison of random sampling, single-syndrome and multi-syndrome for error estimation with 2000 sets of keys at the QBER of 0.0068 (top), 0.0166 (middle), and 0.0267 (bottom), respectively.
(a) Results of multi-syndrome error estimation (black lines) and random sampling method (magenta lines).
(b) Results of multi-syndrome error estimation (black lines) and single-syndrome error estimation (red lines).}
	\label{fig:errorEstimate}
\end{figure*}

We have described the three methods of error estimation hereinbefore, including the random sampling, the single-syndrome error estimation and the multi-syndrome error estimation. To compare these three methods, we generate 2000 sets of keys at error rates of $0.0068$, $0.0166$, and $0.0267$, respectively. The sampling rate of random sampling is set to 0.5. For any set of key, we use these methods to get three error rates. As shown in Fig. \ref{fig:errorEstimate}, it is clear that our method (black lines) is more  accurate and stable than the random sampling (magenta lines) and the single-syndrome error estimation (red lines).

\subsection{Convergence Speed}

\begin{figure}[htbp]
	\centering
    \includegraphics[width=0.48\textwidth]{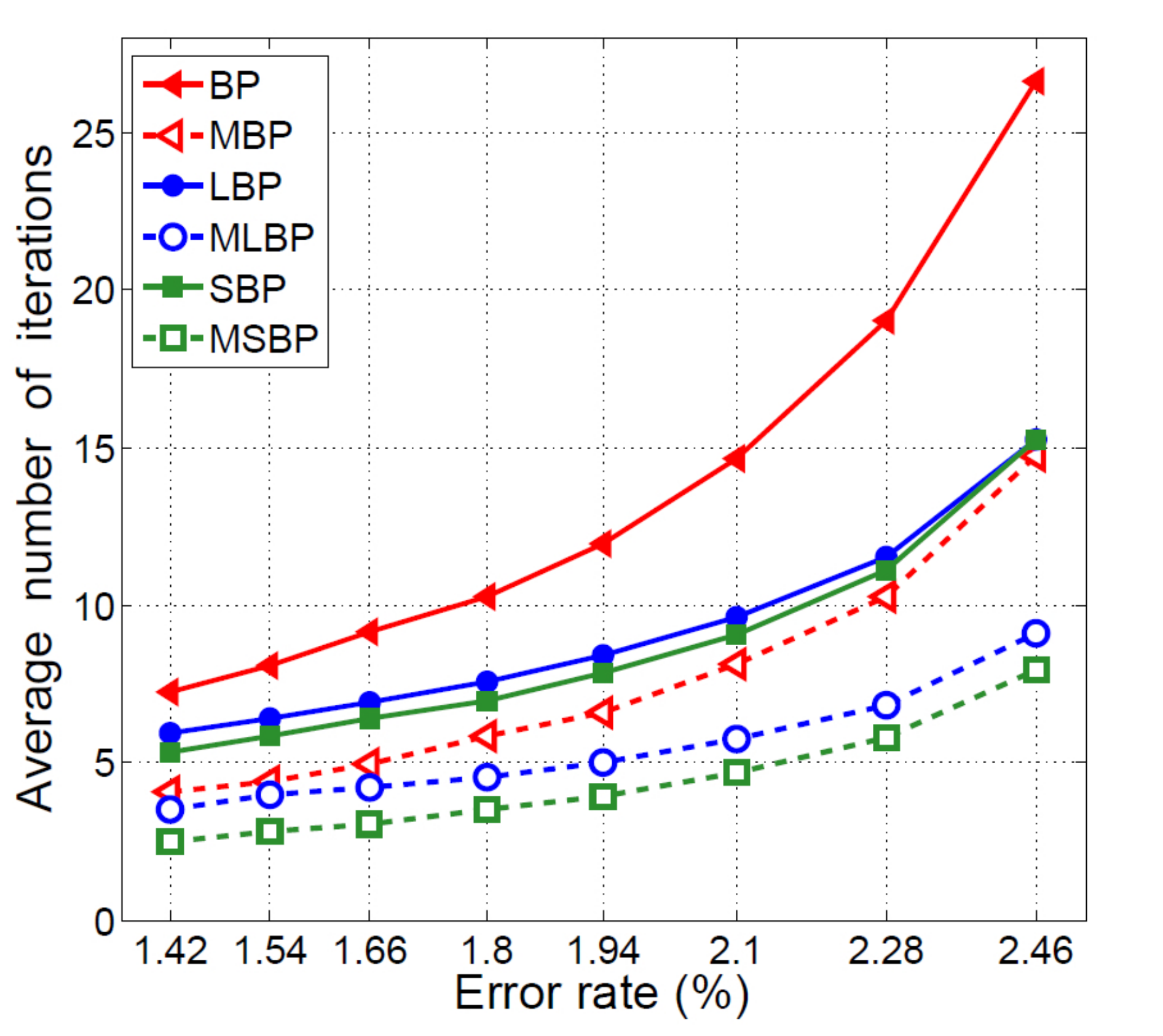}
	\caption{Comparison about convergence speed of 6 reconciliation algorithms by calculating their average numbers of iterations for different error rates.}
	\label{fig:Iteration}
\end{figure}

For key reconciliation, since the faster the convergence speed is, the smaller the average number of iterations becomes,
we evaluate the convergence speed of different algorithms by calculating their average numbers of iterations under different error rates.
We first prepare a matrix for the single-matrix algorithms, then add four more matrices for the multi-matrix algorithms (see the next section for the detailed method of generating LDPC codes). At a certain error rate, we generate 100 sets of keys, perform each algorithm on the keys, and calculate the average number of iterations.
The results are shown in Fig.~\ref{fig:Iteration}. Clearly, under different error rates, the average numbers of iterations of the multi-matrix algorithms are significantly decreased compared with their single-matrix versions. MBP cuts down 43.15$\sim$46.06\% of average iteration number of BP, while MLBP is 38.16$\sim$40.21\% and MSBP is 47.87$\sim$53.38\%.

\begin{figure}[htbp]
	\centering
    \includegraphics[width=0.46\textwidth,height=0.42\textwidth]{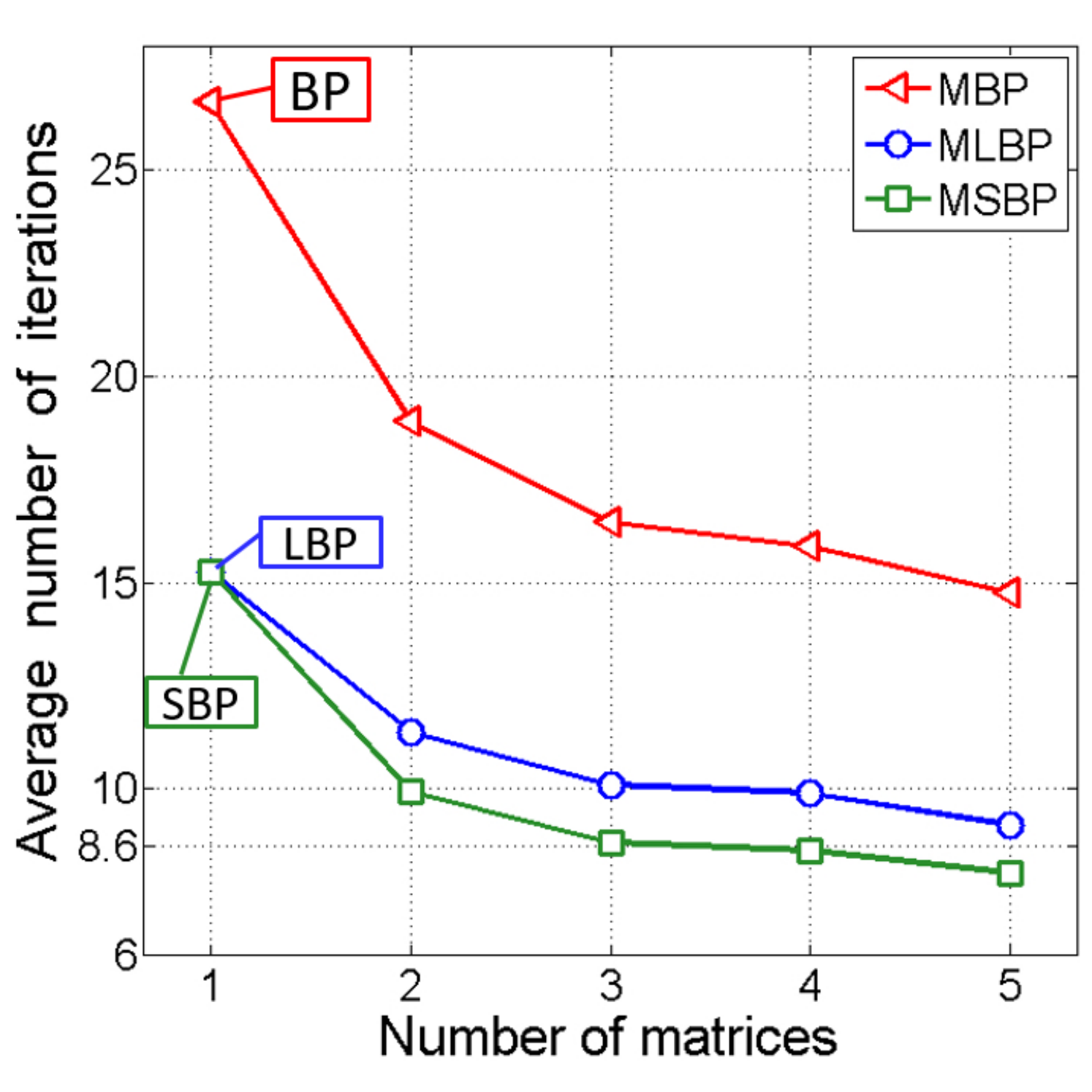}
	\caption{ Relationship about the convergence speed and the number of matrices (1$\sim$5) in reconciliation.
The error rate for data simulation is 0.0246.
}
	\label{fig:Number}
\end{figure}

We can further increase the convergence speed of the multi-matrix algorithms by adjusting two factors. One is the number of matrices used in reconciliation. We generate 100 sets of keys with error rate 0.0246, run the multi-matrix algorithms with different number of matrices to correct these keys. The relationship between the average number of iterations and the number of matrices is plotted in Fig. \ref{fig:Number}. Clearly, the average number of iterations and the number of matrices are inversely proportional.

\begin{figure}[htbp]
	\centering
    \includegraphics[width=0.46\textwidth]{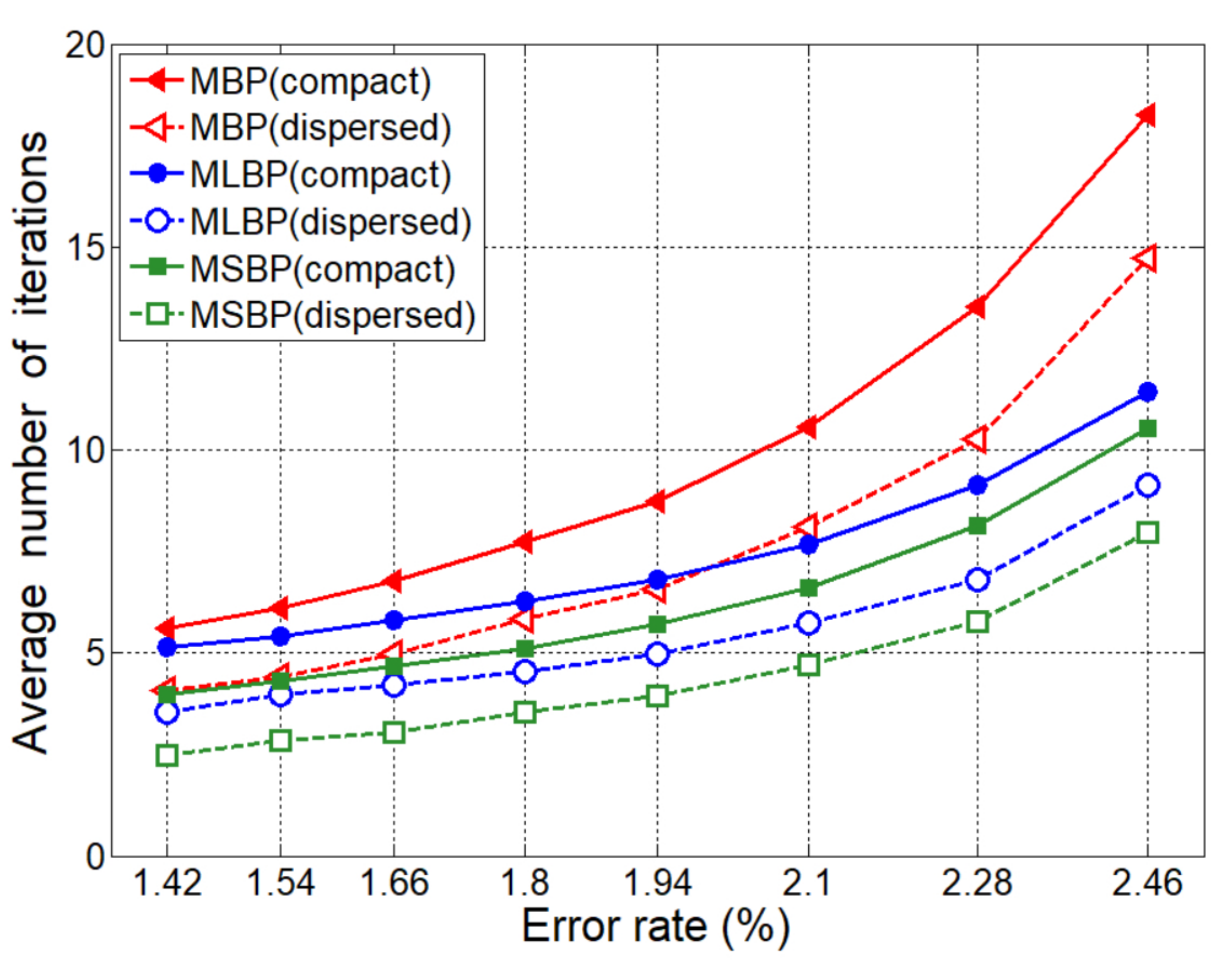}
	\caption{The convergence speed of the multi-matrix algorithms relative to the number of waves is shown.
We generate 100 sets of keys at each error rate, perform each algorithm on the keys using 5 matrices with compact and separated waves  respectively,
and calculate the average number of iterations.}
	\label{fig:Wave}
\end{figure}

Another factor is the number of waves.
The variable nodes with larger degrees can get more information, thus can be corrected earlier and can provide useful information to help other variable nodes.
This process spreads from large-degree to small-degree variable nodes, behaving like a wave, so it is called the wave effect \cite{Luby2001Improved}.
For a multi-matrix algorithm, the multiple waves can be formed simultaneously to correct errors. We refer this phenomenon as the multi-wave effect, which obviously leads to faster convergence speed.
However, if the waves are close to each other, they spread as one wave. This greatly discounts the performance of the multi-wave effect.
On the contrary, if the large-degree variable nodes are dispersed in different matrices, the multiple waves spread and correct errors at the same time, resulting in faster convergence speed.
We construct $5$ matrices with close waves to compare with $5$ matrices with separated ones, and plot the results in Fig.~\ref{fig:Wave}. Clearly, the algorithms using matrices with separated waves outperform the others.

Therefore, our strategy can significantly improve the convergence speed compared with the single-matrix reconciliation, and the speed can be further improved if Bob uses more or designs better matrices.

\subsection{Success Rate}

\begin{figure}[htbp]
	\centering
    \includegraphics[width=0.46\textwidth]{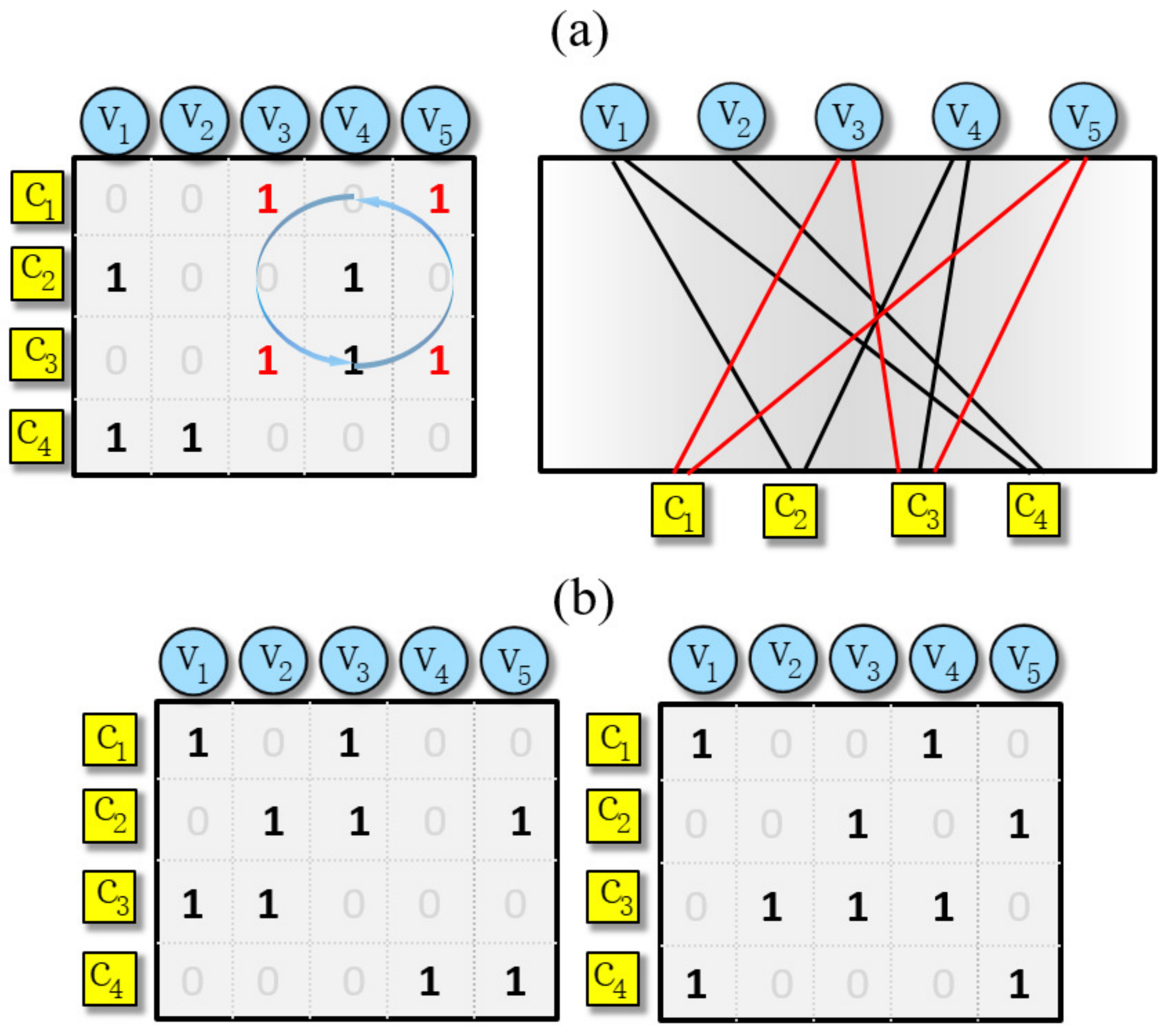}
	\caption{ One matrix with a 4-member cycle (a) and two additional matrices (b). }
	\label{fig:cycle}
\end{figure}

The success rate of reconciliation may be negatively impacted by the cycles.
For example, suppose Alice's sifted key is $x^{T}=[1,0,1,0,1]$, Bob's sifted key is $y^{T}=[1,0,0,0,1]$, the error rate $e$ is 0.2, LDPC code has $5$ variable nodes labeled as $\{v_1,\dots,v_5\}$ and $4$ check nodes denoted as $\{c_1,\dots,c_4\}$. As shown in Fig. \ref{fig:cycle} (a), in LDPC code there is a 4-member cycle which is represented by a blue circle and red edges, respectively. If Bob uses BP algorithm to correct the key, the reconciliation is failed in each iteration. It is because that there is always a difference between the signs of soft-decision values of $v_2$ and $v_4$. Therefore, they cannot be decoded as $1$ at the same time. The 4-member cycle makes new information always be excluded and old information always loop in the cycle. Thus,  as recorded in Tab. \ref{tab:ring-assoliation}, no matter how large the upper limit of iterations is, the single-matrix reconciliation always fails.

However, as shown in Fig.~\ref{fig:cycle} (b), if Bob adds two matrices to correct the key, since there are no cycle between $v_2$ and $v_4$ in the new matrices, the data of the new matrices help $v_2$ and $v_4$ break out of the 4-member cycle, resulting in a successful reconciliation. As shown in Tab. \ref{tab:ring-multi}, MBP correct the error within two iterations.

\begin{table}[htbp]
	\centering
	\caption{\bf Soft-decision Values of $\mathbf{v_2}$ and $\mathbf{v_4}$}
	\begin{tabular}{|p{1.2cm}|p{2.6cm}|p{2.6cm}|p{1.2cm}|}
		\hline
		Iteration number & Soft-decision value of $v_2$ & Soft-decision value of $v_4$ & Result \\
		\hline
		1 & -0.753772 & 0.753772 & fail \\
		2 & 0.753772 & -0.753772 & fail \\
		3 & -0.728434 & 0.728434 & fail \\
		4 & -0.728434 & 0.728434 & fail \\
		5 & -0.704088 & 0.704088 & fail \\
		$\cdots$ & $\cdots$ & $\cdots$ & $\cdots$ \\
		96 & 0.166115 & -0.166115 & fail \\
		97 & -0.160981 & 0.160981 & fail \\
		98 & 0.160981 & -0.160981 & fail \\
		99 & -0.156007 & 0.156007 & fail \\
		100 & 0.156007 & -0.156007 & fail \\
		\hline
	\end{tabular}
	\label{tab:ring-assoliation}
\end{table}

\begin{table*}[htbp]
	\centering
	\caption{\bf Soft-decision Values of $\mathbf{v_2}$ and $\mathbf{v_4}$ in 3-matrix reconciliation}
	\begin{tabular}{|p{1.2cm}|p{1.7cm}|p{1.7cm}|p{1.7cm}|p{1.7cm}|p{1.7cm}|p{1.7cm}|p{1.7cm}|p{1.7cm}|p{1.2cm}|}
		\hline
		Iteration number & Soft-decision value of $v_2$ in $H_1$ & Soft-decision value of $v_2$ in $H_2$ & Soft-decision value of $v_2$ in $H_3$ & Soft-decision value of $v_2$ & Soft-decision value of $v_4$ in $H_1$ & Soft-decision value of $v_4$ in $H_2$ & Soft-decision value of $v_4$ in $H_3$ & Soft-decision value of $v_4$ & Result \\
		\hline
		1 & -0.753772 & -0.753772 & -0.753772 & -5.0339 & 0.753772 & -2.01882 & -1.38629 & 0.121249 & fail \\
		2 & 0.753772 & -3.46963 & -2.56496 & -8.0534 & -0.753772 & -2.77259 & -3.52636 & -4.28013 & success \\
		\hline
	\end{tabular}
	\label{tab:ring-multi}
\end{table*}

\begin{figure}[htbp]
	\centering
    \includegraphics[width=0.46\textwidth]{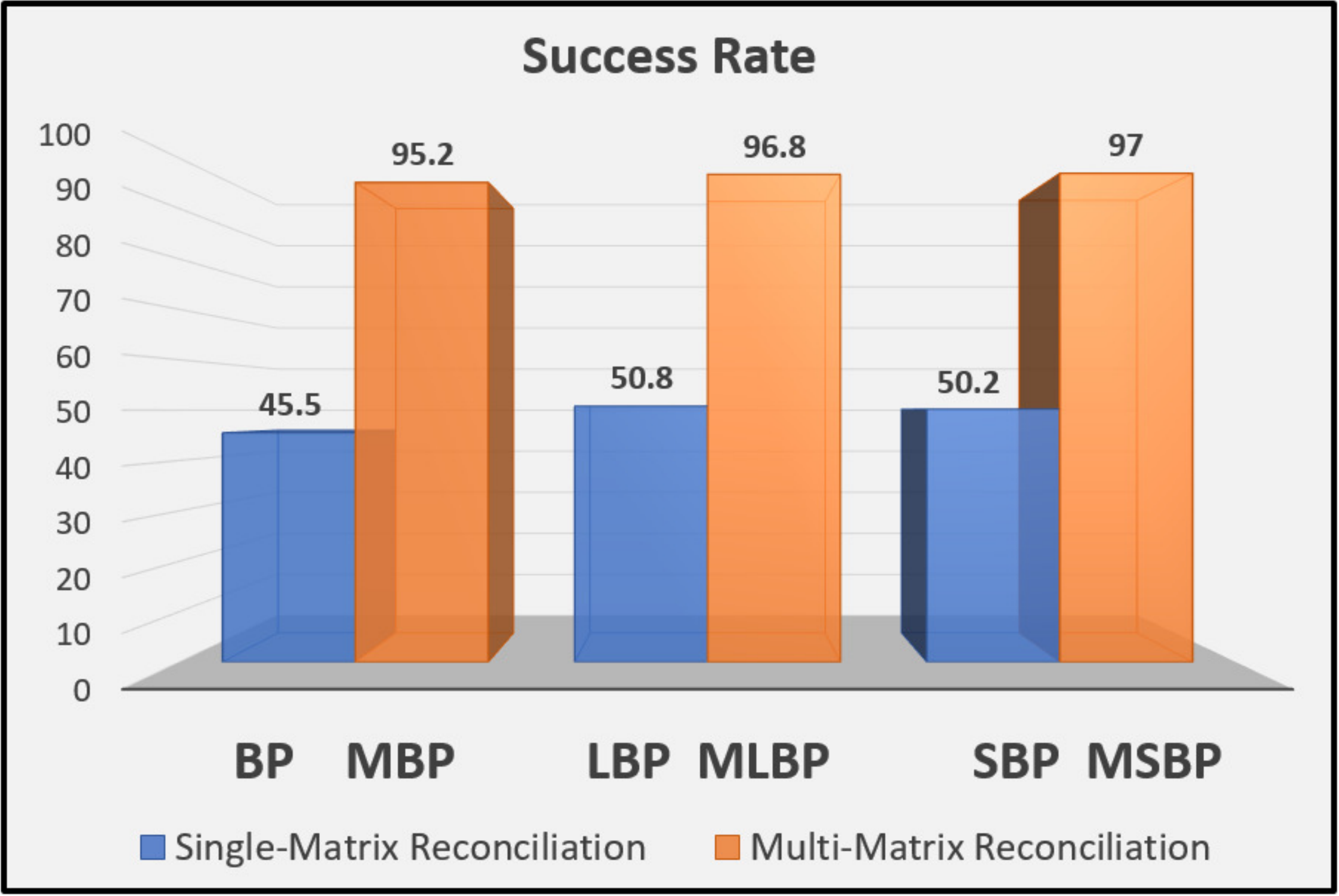}
	\caption{Reconciliation success rate for single- and multi-matrix algorithms.
 1000 sets of keys with error rate of 0.0275 are generated for the comparison.}
	\label{fig:success}
\end{figure}

We carry out a test to fully represent the performance of reducing the impact of cycles.
In this test, we generate 1000 sets of keys with error rate 0.0275, perform the 6 reconciliation algorithms on the generated keys, and calculate the success rate. As shown in Fig. \ref{fig:success}, the average success rate of the multi-matrix algorithms is 96.33\%, nearly double that, 48.83\%, of the single-matrix algorithms.

\subsection{Bit Error Rate}

\begin{figure}[htbp]
	\centering
    \includegraphics[width=0.46\textwidth]{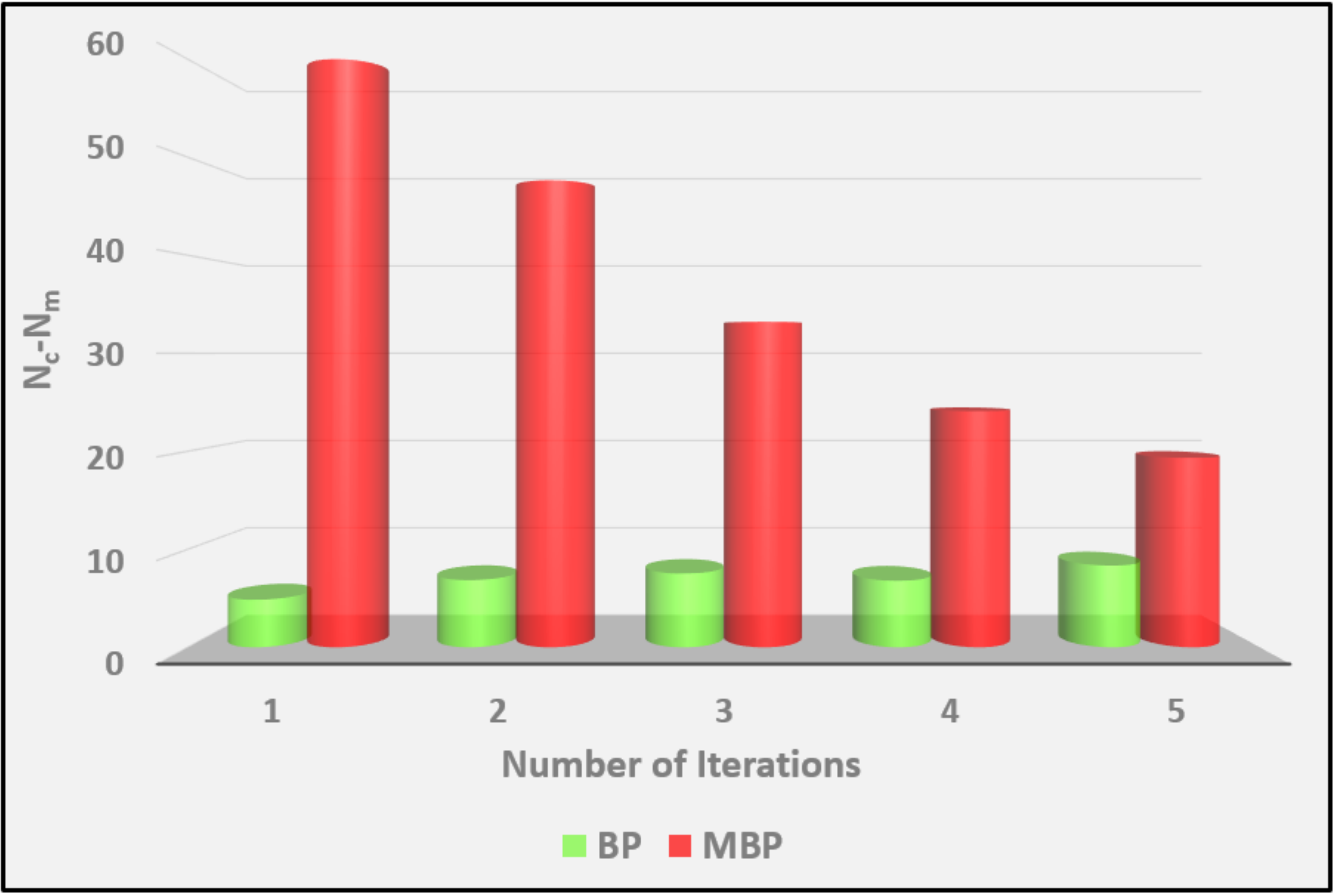}
	\caption{The valid number of corrected bits $N_c-N_m$ ($N_c$ the number of corrected bits; $N_m$ is the number of  misjudged  bits) in each iteration for single- and multi-matrix algorithms. 100 sets of keys with error rate of 0.0267 are considered.}
	\label{fig:right_wrong}
\end{figure}

Compared with the single-matrix reconciliation, the multi-matrix algorithms decode the key according to information provided by multiple matrices. The decoding results are more accurate and reliable. We generate 100 sets of keys with error rate 0.0267, perform BP and MBP on the generated keys to calculate the number of corrected bits $N_c$ and the number of  misjudged  bits $N_m$ in each iteration, and plot the valid number of corrected bits $N_c-N_m$ in Fig. \ref{fig:right_wrong}.
We can see that MBP can correct more errors in each iteration, and most of the errors are corrected at the beginnings of the iterations. It achieves faster convergence speed and lower BER compared with BP.

\begin{figure}[htbp]
	\centering
    \includegraphics[width=0.46\textwidth,height=0.42\textwidth]{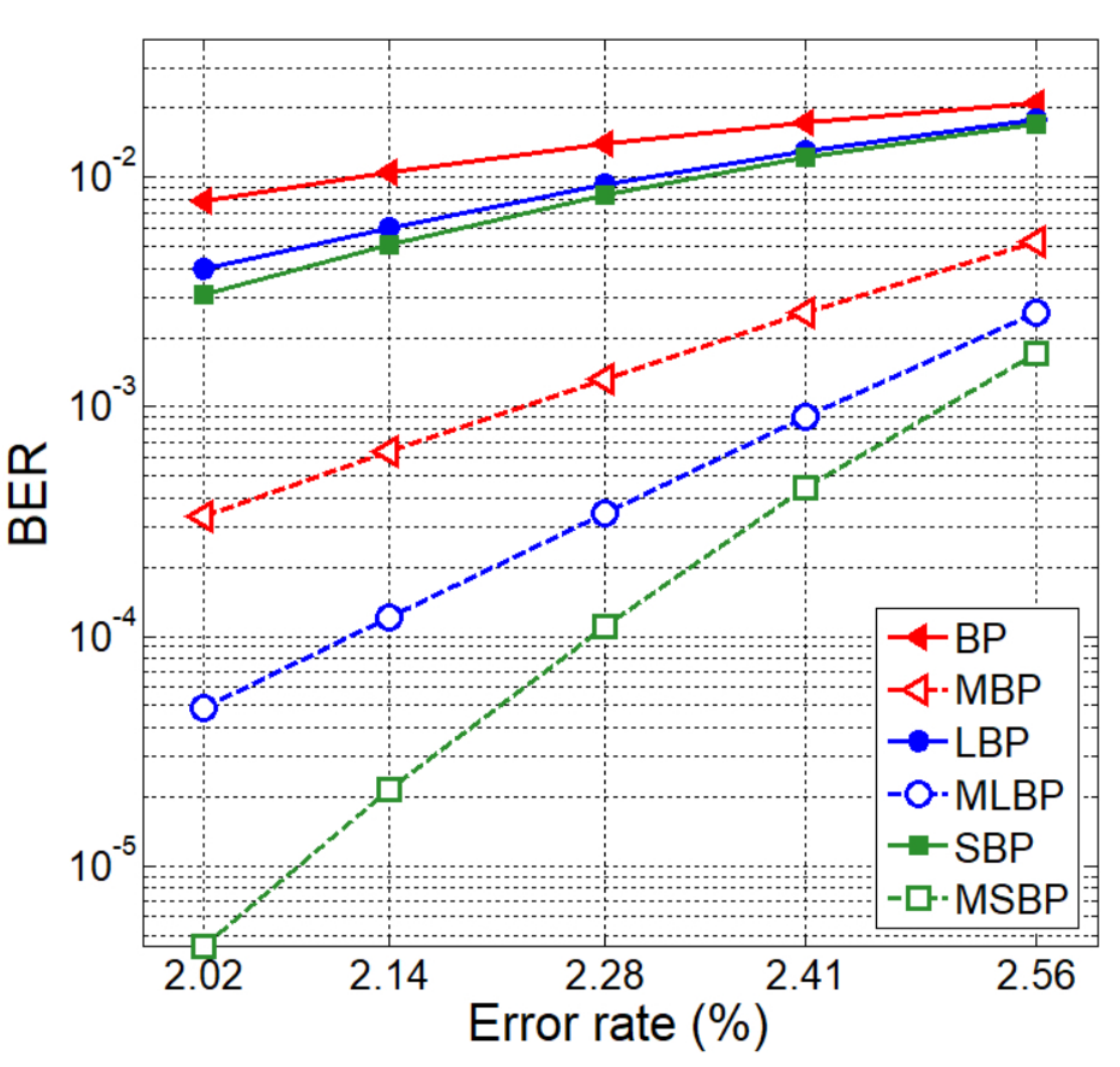}
	\caption{The BER performances of the multi-matrix algorithms after 5 iterations are shown. Five QBER ranging from 0.0202 to 0.0256 are selected. For each error rate, we generate 1000 sets of keys, perform 5-matrix algorithms and single-matrix versions on these generated keys.}
	\label{fig:BER}
\end{figure}

To further evaluate the BER performances of the multi-matrix algorithms, five QBER values ranging from 0.0202 to 0.0256 are selected. At each error rate, we generate 1000 sets of keys, perform 5-matrix algorithms and their single-matrix versions on these generated keys. After 5 iterations, we calculate BERs of different algorithms according to the following equation,

\begin{equation}
\mathrm{BER}=\frac{\mathrm{number\ of\ error\  bits}}{\mathrm{1000 * length\ of\ code}},
\label{equ:BER}
\end{equation}

\noindent and draw the results in Fig. \ref{fig:BER}. It is obvious that all three multi-matrix algorithms achieve lower BERs under different error rates compared with their single-matrix versions.
For example, the BER of SBP is 0.0030832 when the error rate is 0.0202, while MSBP is 0.0000045, between which there is a difference of 3-order magnitude.

\section{Conclusion}

In this paper, a highly efficient error reconciliation protocol for QKD is proposed, whose core is using likelihood of multiple syndromes
obtained from multiple LDPC codes for QBER estimation and correction. Security analysis and multi-matrix construction method are provided.
Evaluation results show that
the proposed approach allows improving the accuracy of QBER estimation in contract to previous works. Additionally, the scheme
can greatly increase the convergence speed, success rate, and significantly improve the BER performance during key reconciliation without compromising
the reconciliation efficiency and significant expenditure of authentication and time resources. Our
findings can lower the complexity for post-processing procedure, thus will promote the commercialization
of QKD.

\section{Acknowledgements}
This research is financially supported by the Major Program of National Natural Science Foundation of China (No. 11690030, 11690032, 11804153), the National Key Research and Development Program of China (No. 2017YFA0303700), the National Natural Science Foundation of China (No. 61771236), and the Excellence Research Program of Nanjing University.
The authors are grateful to Wen-yuan Wang for valuable contribution.

\bibliography{apssamp}

\begin{thebibliography}{37}%
\makeatletter
\providecommand \@ifxundefined [1]{%
 \@ifx{#1\undefined}
}%
\providecommand \@ifnum [1]{%
 \ifnum #1\expandafter \@firstoftwo
 \else \expandafter \@secondoftwo
 \fi
}%
\providecommand \@ifx [1]{%
 \ifx #1\expandafter \@firstoftwo
 \else \expandafter \@secondoftwo
 \fi
}%
\providecommand \natexlab [1]{#1}%
\providecommand \enquote  [1]{``#1''}%
\providecommand \bibnamefont  [1]{#1}%
\providecommand \bibfnamefont [1]{#1}%
\providecommand \citenamefont [1]{#1}%
\providecommand \href@noop [0]{\@secondoftwo}%
\providecommand \href [0]{\begingroup \@sanitize@url \@href}%
\providecommand \@href[1]{\@@startlink{#1}\@@href}%
\providecommand \@@href[1]{\endgroup#1\@@endlink}%
\providecommand \@sanitize@url [0]{\catcode `\\12\catcode `\$12\catcode
  `\&12\catcode `\#12\catcode `\^12\catcode `\_12\catcode `\%12\relax}%
\providecommand \@@startlink[1]{}%
\providecommand \@@endlink[0]{}%
\providecommand \url  [0]{\begingroup\@sanitize@url \@url }%
\providecommand \@url [1]{\endgroup\@href {#1}{\urlprefix }}%
\providecommand \urlprefix  [0]{URL }%
\providecommand \Eprint [0]{\href }%
\providecommand \doibase [0]{http://dx.doi.org/}%
\providecommand \selectlanguage [0]{\@gobble}%
\providecommand \bibinfo  [0]{\@secondoftwo}%
\providecommand \bibfield  [0]{\@secondoftwo}%
\providecommand \translation [1]{[#1]}%
\providecommand \BibitemOpen [0]{}%
\providecommand \bibitemStop [0]{}%
\providecommand \bibitemNoStop [0]{.\EOS\space}%
\providecommand \EOS [0]{\spacefactor3000\relax}%
\providecommand \BibitemShut  [1]{\csname bibitem#1\endcsname}%
\let\auto@bib@innerbib\@empty
\bibitem [{\citenamefont {Gisin}\ \emph {et~al.}(2002)\citenamefont {Gisin},
  \citenamefont {Ribordy}, \citenamefont {Tittel},\ and\ \citenamefont
  {Zbinden}}]{gisin2002quantum}%
  \BibitemOpen
  \bibfield  {author} {\bibinfo {author} {\bibfnamefont {N.}~\bibnamefont
  {Gisin}}, \bibinfo {author} {\bibfnamefont {G.}~\bibnamefont {Ribordy}},
  \bibinfo {author} {\bibfnamefont {W.}~\bibnamefont {Tittel}}, \ and\ \bibinfo
  {author} {\bibfnamefont {H.}~\bibnamefont {Zbinden}},\ }\bibfield  {title}
  {\bibinfo {title} {Quantum cryptography},\ }\href@noop {} {\bibfield
  {journal} {\bibinfo  {journal}
  {\href{https://journals.aps.org/rmp/abstract/10.1103/RevModPhys.74.145}{Rev.
  Mod. Phys. }}}\textbf {\bibinfo {volume}
  {\href{https://journals.aps.org/rmp/abstract/10.1103/RevModPhys.74.145}{74}}}\bibinfo
  {pages}
  {\href{https://journals.aps.org/rmp/abstract/10.1103/RevModPhys.74.145}{,
  145}}\bibinfo {year}
  {\href{https://journals.aps.org/rmp/abstract/10.1103/RevModPhys.74.145}{
  (2002).}}}\BibitemShut {Stop}%
\bibitem [{\citenamefont {Scarani}\ \emph {et~al.}(2009)\citenamefont
  {Scarani}, \citenamefont {Bechmann-Pasquinucci}, \citenamefont {Cerf},
  \citenamefont {Du{\v{s}}ek}, \citenamefont {L{\"u}tkenhaus},\ and\
  \citenamefont {Peev}}]{scarani2009security}%
  \BibitemOpen
  \bibfield  {author} {\bibinfo {author} {\bibfnamefont {V.}~\bibnamefont
  {Scarani}}, \bibinfo {author} {\bibfnamefont {H.}~\bibnamefont
  {Bechmann-Pasquinucci}}, \bibinfo {author} {\bibfnamefont {N.~J.}\
  \bibnamefont {Cerf}}, \bibinfo {author} {\bibfnamefont {M.}~\bibnamefont
  {Du{\v{s}}ek}}, \bibinfo {author} {\bibfnamefont {N.}~\bibnamefont
  {L{\"u}tkenhaus}}, \ and\ \bibinfo {author} {\bibfnamefont {M.}~\bibnamefont
  {Peev}},\ }\bibfield  {title} {\bibinfo {title} {The security of practical
  quantum key distribution},\ }\href@noop {} {\bibfield  {journal} {\bibinfo
  {journal}
  {\href{https://journals.aps.org/rmp/abstract/10.1103/RevModPhys.81.1301}{Rev.
  Mod. Phys. }}}\textbf {\bibinfo {volume}
  {\href{https://journals.aps.org/rmp/abstract/10.1103/RevModPhys.81.1301}{81}}}\bibinfo
  {pages}
  {\href{https://journals.aps.org/rmp/abstract/10.1103/RevModPhys.81.1301}{,
  1301}}\bibinfo {year}
  {\href{https://journals.aps.org/rmp/abstract/10.1103/RevModPhys.81.1301}{
  (2009).}}}\BibitemShut {Stop}%
\bibitem [{\citenamefont {Lo}\ \emph {et~al.}(2005)\citenamefont {Lo},
  \citenamefont {Ma},\ and\ \citenamefont {Chen}}]{lo2005decoy}%
  \BibitemOpen
  \bibfield  {author} {\bibinfo {author} {\bibfnamefont {H.-K.}\ \bibnamefont
  {Lo}}, \bibinfo {author} {\bibfnamefont {X.}~\bibnamefont {Ma}}, \ and\
  \bibinfo {author} {\bibfnamefont {K.}~\bibnamefont {Chen}},\ }\bibfield
  {title} {\bibinfo {title} {Decoy state quantum key distribution},\
  }\href@noop {} {\bibfield  {journal} {\bibinfo  {journal}
  {\href{https://journals.aps.org/prl/abstract/10.1103/PhysRevLett.94.230504}{Phys.
  Rev. Lett. }}}\textbf {\bibinfo {volume}
  {\href{https://journals.aps.org/prl/abstract/10.1103/PhysRevLett.94.230504}{94}}}\bibinfo
  {pages}
  {\href{https://journals.aps.org/prl/abstract/10.1103/PhysRevLett.94.230504}{,
  230504}}\bibinfo {year}
  {\href{https://journals.aps.org/prl/abstract/10.1103/PhysRevLett.94.230504}{
  (2005).}}}\BibitemShut {Stop}%
\bibitem [{\citenamefont {Lo}\ \emph {et~al.}(2012)\citenamefont {Lo},
  \citenamefont {Curty},\ and\ \citenamefont {Qi}}]{lo2012measurement}%
  \BibitemOpen
  \bibfield  {author} {\bibinfo {author} {\bibfnamefont {H.-K.}\ \bibnamefont
  {Lo}}, \bibinfo {author} {\bibfnamefont {M.}~\bibnamefont {Curty}}, \ and\
  \bibinfo {author} {\bibfnamefont {B.}~\bibnamefont {Qi}},\ }\bibfield
  {title} {\bibinfo {title} {Measurement-device-independent quantum key
  distribution},\ }\href@noop {} {\bibfield  {journal} {\bibinfo  {journal}
  {\href{https://journals.aps.org/prl/abstract/10.1103/PhysRevLett.108.130503}{Phys.
  Rev. Lett. }}}\textbf {\bibinfo {volume}
  {\href{https://journals.aps.org/prl/abstract/10.1103/PhysRevLett.108.130503}{108}}}\bibinfo
  {pages}
  {\href{https://journals.aps.org/prl/abstract/10.1103/PhysRevLett.108.130503}{,
  130503}}\bibinfo {year}
  {\href{https://journals.aps.org/prl/abstract/10.1103/PhysRevLett.108.130503}{
  (2012).}}}\BibitemShut {Stop}%
\bibitem [{\citenamefont {Liao}\ \emph {et~al.}(2017)\citenamefont {Liao},
  \citenamefont {Cai}, \citenamefont {Liu}, \citenamefont {Zhang},
  \citenamefont {Li}, \citenamefont {Ren}, \citenamefont {Yin}, \citenamefont
  {Shen}, \citenamefont {Cao}, \citenamefont {Li} \emph
  {et~al.}}]{liao2017satellite}%
  \BibitemOpen
  \bibfield  {author} {\bibinfo {author} {\bibfnamefont {S.-K.}\ \bibnamefont
  {Liao}}, \bibinfo {author} {\bibfnamefont {W.-Q.}\ \bibnamefont {Cai}},
  \bibinfo {author} {\bibfnamefont {W.-Y.}\ \bibnamefont {Liu}}, \bibinfo
  {author} {\bibfnamefont {L.}~\bibnamefont {Zhang}}, \bibinfo {author}
  {\bibfnamefont {Y.}~\bibnamefont {Li}}, \bibinfo {author} {\bibfnamefont
  {J.-G.}\ \bibnamefont {Ren}}, \bibinfo {author} {\bibfnamefont
  {J.}~\bibnamefont {Yin}}, \bibinfo {author} {\bibfnamefont {Q.}~\bibnamefont
  {Shen}}, \bibinfo {author} {\bibfnamefont {Y.}~\bibnamefont {Cao}}, \bibinfo
  {author} {\bibfnamefont {Z.-P.}\ \bibnamefont {Li}},  \emph {et~al.},\
  }\bibfield  {title} {\bibinfo {title} {Satellite-to-ground quantum key
  distribution},\ }\href@noop {} {\bibfield  {journal} {\bibinfo  {journal}
  {\href{https://www.nature.com/articles/nature23655}{Nature }}}\textbf
  {\bibinfo {volume}
  {\href{https://www.nature.com/articles/nature23655}{549}}}\bibinfo {pages}
  {\href{https://www.nature.com/articles/nature23655}{, 43}}\bibinfo {year}
  {\href{https://www.nature.com/articles/nature23655}{ (2017).}}}\BibitemShut
  {Stop}%
\bibitem [{\citenamefont {Wang}(2005)}]{Wang2005Beating}%
  \BibitemOpen
  \bibfield  {author} {\bibinfo {author} {\bibfnamefont {X.~B.}\ \bibnamefont
  {Wang}},\ }\bibfield  {title} {\bibinfo {title} {Beating the
  photon-number-splitting attack in practical quantum cryptography},\
  }\href@noop {} {\bibfield  {journal} {\bibinfo  {journal}
  {\href{https://journals.aps.org/prl/abstract/10.1103/PhysRevLett.94.230503}{Phys.
  Rev. Lett. }}}\textbf {\bibinfo {volume}
  {\href{https://journals.aps.org/prl/abstract/10.1103/PhysRevLett.94.230503}{94}}}\bibinfo
  {pages}
  {\href{https://journals.aps.org/prl/abstract/10.1103/PhysRevLett.94.230503}{,
  230503}}\bibinfo {year}
  {\href{https://journals.aps.org/prl/abstract/10.1103/PhysRevLett.94.230503}{
  (2005).}}}\BibitemShut {Stop}%
\bibitem [{\citenamefont {Bennet}\ and\ \citenamefont
  {Brassard}(1984)}]{Bennet1984Quantum}%
  \BibitemOpen
  \bibfield  {author} {\bibinfo {author} {\bibfnamefont {C.~H.}\ \bibnamefont
  {Bennet}}\ and\ \bibinfo {author} {\bibfnamefont {G.}~\bibnamefont
  {Brassard}},\ }in\ \href@noop {} {\emph {\bibinfo {booktitle} {1984 IEEE
  International Conference on Computers, Systems, and Signal processing,
  Bangalore, India}}}\ (\bibinfo {year} {IEEE, 1984}),\ p.\ \bibinfo {pages}
  {175}\BibitemShut {NoStop}%
\bibitem [{\citenamefont {Gerhardt}\ \emph {et~al.}(2010)\citenamefont
  {Gerhardt}, \citenamefont {Liu}, \citenamefont {Lamas-Linares}, \citenamefont
  {Skaar}, \citenamefont {Kurtsiefer},\ and\ \citenamefont
  {Makarov}}]{Gerhardt2010Full}%
  \BibitemOpen
  \bibfield  {author} {\bibinfo {author} {\bibfnamefont {I.}~\bibnamefont
  {Gerhardt}}, \bibinfo {author} {\bibfnamefont {Q.}~\bibnamefont {Liu}},
  \bibinfo {author} {\bibfnamefont {A.}~\bibnamefont {Lamas-Linares}}, \bibinfo
  {author} {\bibfnamefont {J.}~\bibnamefont {Skaar}}, \bibinfo {author}
  {\bibfnamefont {C.}~\bibnamefont {Kurtsiefer}}, \ and\ \bibinfo {author}
  {\bibfnamefont {V.}~\bibnamefont {Makarov}},\ }\bibfield  {title} {\bibinfo
  {title} {Full-field implementation of a perfect eavesdropper on a quantum
  cryptography system},\ }\href@noop {} {\bibfield  {journal} {\bibinfo
  {journal} {\href{https://www.nature.com/articles/ncomms1348}{Nat. Commun.
  }}}\textbf {\bibinfo {volume}
  {\href{https://www.nature.com/articles/ncomms1348}{2}}}\bibinfo {pages}
  {\href{https://www.nature.com/articles/ncomms1348}{, 349}}\bibinfo {year}
  {\href{https://www.nature.com/articles/ncomms1348}{ (2010).}}}\BibitemShut
  {Stop}%
\bibitem [{\citenamefont {Weier}\ \emph {et~al.}(2011)\citenamefont {Weier},
  \citenamefont {Krauss}, \citenamefont {Rau}, \citenamefont {Fuerst},
  \citenamefont {Nauerth},\ and\ \citenamefont
  {Weinfurter}}]{Weier2011Quantum}%
  \BibitemOpen
  \bibfield  {author} {\bibinfo {author} {\bibfnamefont {H.}~\bibnamefont
  {Weier}}, \bibinfo {author} {\bibfnamefont {H.}~\bibnamefont {Krauss}},
  \bibinfo {author} {\bibfnamefont {M.}~\bibnamefont {Rau}}, \bibinfo {author}
  {\bibfnamefont {M.}~\bibnamefont {Fuerst}}, \bibinfo {author} {\bibfnamefont
  {S.}~\bibnamefont {Nauerth}}, \ and\ \bibinfo {author} {\bibfnamefont
  {H.}~\bibnamefont {Weinfurter}},\ }\bibfield  {title} {\bibinfo {title}
  {Quantum eavesdropping without interception: An attack exploiting the dead
  time of single photon detectors},\ }\href@noop {} {\bibfield  {journal}
  {\bibinfo  {journal}
  {\href{https://inis.iaea.org/search/search.aspx?orig_q=RN:43028953}{New. J.
  Phys. }}}\textbf {\bibinfo {volume}
  {\href{https://inis.iaea.org/search/search.aspx?orig_q=RN:43028953}{13}}}\bibinfo
  {pages} {\href{https://inis.iaea.org/search/search.aspx?orig_q=RN:43028953}{,
  193}}\bibinfo {year}
  {\href{https://inis.iaea.org/search/search.aspx?orig_q=RN:43028953}{
  (2011).}}}\BibitemShut {Stop}%
\bibitem [{\citenamefont {Jain}\ \emph {et~al.}(2011)\citenamefont {Jain},
  \citenamefont {Wittmann}, \citenamefont {Lydersen}, \citenamefont {Wiechers},
  \citenamefont {Elser}, \citenamefont {Marquardt}, \citenamefont {Makarov},\
  and\ \citenamefont {Leuchs}}]{Jain2011Device}%
  \BibitemOpen
  \bibfield  {author} {\bibinfo {author} {\bibfnamefont {N.}~\bibnamefont
  {Jain}}, \bibinfo {author} {\bibfnamefont {C.}~\bibnamefont {Wittmann}},
  \bibinfo {author} {\bibfnamefont {L.}~\bibnamefont {Lydersen}}, \bibinfo
  {author} {\bibfnamefont {C.}~\bibnamefont {Wiechers}}, \bibinfo {author}
  {\bibfnamefont {D.}~\bibnamefont {Elser}}, \bibinfo {author} {\bibfnamefont
  {C.}~\bibnamefont {Marquardt}}, \bibinfo {author} {\bibfnamefont
  {V.}~\bibnamefont {Makarov}}, \ and\ \bibinfo {author} {\bibfnamefont
  {G.}~\bibnamefont {Leuchs}},\ }\bibfield  {title} {\bibinfo {title} {Device
  calibration impacts security of quantum key distribution},\ }\href@noop {}
  {\bibfield  {journal} {\bibinfo  {journal}
  {\href{https://journals.aps.org/prl/abstract/10.1103/PhysRevLett.107.110501}{Phys.
  Rev. Lett. }}}\textbf {\bibinfo {volume}
  {\href{https://journals.aps.org/prl/abstract/10.1103/PhysRevLett.107.110501}{107}}}\bibinfo
  {pages}
  {\href{https://journals.aps.org/prl/abstract/10.1103/PhysRevLett.107.110501}{,
  110501}}\bibinfo {year}
  {\href{https://journals.aps.org/prl/abstract/10.1103/PhysRevLett.107.110501}{
  (2011).}}}\BibitemShut {Stop}%
\bibitem [{\citenamefont {Treeviriyanupab}\ \emph {et~al.}(2014)\citenamefont
  {Treeviriyanupab}, \citenamefont {Phromsa-ard}, \citenamefont {Zhang},
  \citenamefont {Li}, \citenamefont {Sangwongngam}, \citenamefont {Ayutaya},
  \citenamefont {Songneam}, \citenamefont {Rattanatamma}, \citenamefont
  {Ingkavet}, \citenamefont {Sanor} \emph {et~al.}}]{treeviriyanupab2014rate}%
  \BibitemOpen
  \bibfield  {author} {\bibinfo {author} {\bibfnamefont {P.}~\bibnamefont
  {Treeviriyanupab}}, \bibinfo {author} {\bibfnamefont {T.}~\bibnamefont
  {Phromsa-ard}}, \bibinfo {author} {\bibfnamefont {C.-M.}\ \bibnamefont
  {Zhang}}, \bibinfo {author} {\bibfnamefont {M.}~\bibnamefont {Li}}, \bibinfo
  {author} {\bibfnamefont {P.}~\bibnamefont {Sangwongngam}}, \bibinfo {author}
  {\bibfnamefont {T.~S.~N.}\ \bibnamefont {Ayutaya}}, \bibinfo {author}
  {\bibfnamefont {N.}~\bibnamefont {Songneam}}, \bibinfo {author}
  {\bibfnamefont {R.}~\bibnamefont {Rattanatamma}}, \bibinfo {author}
  {\bibfnamefont {C.}~\bibnamefont {Ingkavet}}, \bibinfo {author}
  {\bibfnamefont {W.}~\bibnamefont {Sanor}},  \emph {et~al.},\ }in\ \href@noop
  {} {\emph {\bibinfo {booktitle} {2014 14th International Symposium on
  Communications and Information Technologies, Incheon, South Korea}}}\
  (\bibinfo {year} {IEEE, 2014}),\ p.\ \bibinfo {pages} {351}\BibitemShut
  {NoStop}%
\bibitem [{\citenamefont {Kiktenko}\ \emph {et~al.}(2018)\citenamefont
  {Kiktenko}, \citenamefont {Malyshev}, \citenamefont {Bozhedarov},
  \citenamefont {Pozhar}, \citenamefont {Anufriev},\ and\ \citenamefont
  {Fedorov}}]{2018arXiv181005841K}%
  \BibitemOpen
  \bibfield  {author} {\bibinfo {author} {\bibfnamefont {E.}~\bibnamefont
  {Kiktenko}}, \bibinfo {author} {\bibfnamefont {A.}~\bibnamefont {Malyshev}},
  \bibinfo {author} {\bibfnamefont {A.}~\bibnamefont {Bozhedarov}}, \bibinfo
  {author} {\bibfnamefont {N.}~\bibnamefont {Pozhar}}, \bibinfo {author}
  {\bibfnamefont {M.}~\bibnamefont {Anufriev}}, \ and\ \bibinfo {author}
  {\bibfnamefont {A.}~\bibnamefont {Fedorov}},\ }\bibfield  {title} {\bibinfo
  {title} {Error estimation at the information reconciliation stage of quantum
  key distribution},\ }\href@noop {} {\bibfield  {journal} {\bibinfo  {journal}
  {J. Russ. Laser. Res.}}\textbf {\bibinfo {volume} {39}}\bibinfo {pages}
  {558}\bibinfo {year} {2018}}\BibitemShut {NoStop}%
\bibitem [{\citenamefont {Luby}\ \emph {et~al.}(1998)\citenamefont {Luby},
  \citenamefont {Amin~Shokrolloahi}, \citenamefont {Mizenmacher},\ and\
  \citenamefont {Spielman}}]{Luby1998Improved}%
  \BibitemOpen
  \bibfield  {author} {\bibinfo {author} {\bibfnamefont {M.~G.}\ \bibnamefont
  {Luby}}, \bibinfo {author} {\bibfnamefont {M.}~\bibnamefont
  {Amin~Shokrolloahi}}, \bibinfo {author} {\bibfnamefont {M.}~\bibnamefont
  {Mizenmacher}}, \ and\ \bibinfo {author} {\bibfnamefont {D.~A.}\ \bibnamefont
  {Spielman}},\ }in\ \href@noop {} {\emph {\bibinfo {booktitle} {1998 IEEE
  International Symposium on Information Theory, Cambridge, MA, England}}}\
  (\bibinfo {year} {IEEE, 1998}),\ p.\ \bibinfo {pages} {117}\BibitemShut
  {NoStop}%
\bibitem [{\citenamefont {Bennett}\ \emph {et~al.}(1988)\citenamefont
  {Bennett}, \citenamefont {H}, \citenamefont {Brassard}, \citenamefont
  {Gilles}, \citenamefont {Robert},\ and\ \citenamefont
  {JeanMarc}}]{Bennett1988Privacy}%
  \BibitemOpen
  \bibfield  {author} {\bibinfo {author} {\bibnamefont {Bennett}}, \bibinfo
  {author} {\bibfnamefont {C.}~\bibnamefont {H}}, \bibinfo {author}
  {\bibnamefont {Brassard}}, \bibinfo {author} {\bibnamefont {Gilles}},
  \bibinfo {author} {\bibnamefont {Robert}}, \ and\ \bibinfo {author}
  {\bibnamefont {JeanMarc}},\ }\bibfield  {title} {\bibinfo {title} {Privacy
  amplification by public discussion},\ }\href@noop {} {\bibfield  {journal}
  {\bibinfo  {journal}
  {\href{https://epubs.siam.org/doi/abs/10.1137/0217014}{SIAM J. Comput.
  }}}\textbf {\bibinfo {volume}
  {\href{https://epubs.siam.org/doi/abs/10.1137/0217014}{17}}}\bibinfo {pages}
  {\href{https://epubs.siam.org/doi/abs/10.1137/0217014}{, 210}}\bibinfo {year}
  {\href{https://epubs.siam.org/doi/abs/10.1137/0217014}{ (1988)}}}\BibitemShut
  {NoStop}%
\bibitem [{\citenamefont {Bennett}\ \emph {et~al.}(1995)\citenamefont
  {Bennett}, \citenamefont {Brassard}, \citenamefont {Crepeau},\ and\
  \citenamefont {Maurer}}]{Bennett1995Generalized}%
  \BibitemOpen
  \bibfield  {author} {\bibinfo {author} {\bibfnamefont {C.~H.}\ \bibnamefont
  {Bennett}}, \bibinfo {author} {\bibfnamefont {G.}~\bibnamefont {Brassard}},
  \bibinfo {author} {\bibfnamefont {C.}~\bibnamefont {Crepeau}}, \ and\
  \bibinfo {author} {\bibfnamefont {U.~M.}\ \bibnamefont {Maurer}},\ }\bibfield
   {title} {\bibinfo {title} {Generalized privacy amplification},\ }\href@noop
  {} {\bibfield  {journal} {\bibinfo  {journal}
  {\href{https://ieeexplore.ieee.org/abstract/document/476316/}{IEEE Trans.
  Inform. Theory. }}}\textbf {\bibinfo {volume}
  {\href{https://ieeexplore.ieee.org/abstract/document/476316/}{41}}}\bibinfo
  {pages} {\href{https://ieeexplore.ieee.org/abstract/document/476316/}{,
  1915}}\bibinfo {year}
  {\href{https://ieeexplore.ieee.org/abstract/document/476316/}{
  (1995).}}}\BibitemShut {Stop}%
\bibitem [{\citenamefont {Kou}\ \emph {et~al.}(2001)\citenamefont {Kou},
  \citenamefont {Lin},\ and\ \citenamefont {Fossorier}}]{Kou2001Low}%
  \BibitemOpen
  \bibfield  {author} {\bibinfo {author} {\bibfnamefont {Y.}~\bibnamefont
  {Kou}}, \bibinfo {author} {\bibfnamefont {S.}~\bibnamefont {Lin}}, \ and\
  \bibinfo {author} {\bibfnamefont {M.~P.~C.}\ \bibnamefont {Fossorier}},\
  }\bibfield  {title} {\bibinfo {title} {Low-density parity-check codes based
  on finite geometries: a rediscovery and new results},\ }\href@noop {}
  {\bibfield  {journal} {\bibinfo  {journal}
  {\href{https://ieeexplore.ieee.org/abstract/document/959255/}{IEEE Trans.
  Inform. Theory. }}}\textbf {\bibinfo {volume}
  {\href{https://ieeexplore.ieee.org/abstract/document/959255/}{47}}}\bibinfo
  {pages} {\href{https://ieeexplore.ieee.org/abstract/document/959255/}{,
  2711}}\bibinfo {year}
  {\href{https://ieeexplore.ieee.org/abstract/document/959255/}{
  (2001).}}}\BibitemShut {Stop}%
\bibitem [{\citenamefont {Zhang}\ and\ \citenamefont
  {Fossorier}(2002)}]{Zhang2002Shuffled}%
  \BibitemOpen
  \bibfield  {author} {\bibinfo {author} {\bibfnamefont {J.}~\bibnamefont
  {Zhang}}\ and\ \bibinfo {author} {\bibfnamefont {M.}~\bibnamefont
  {Fossorier}},\ }in\ \href@noop {} {\emph {\bibinfo {booktitle} {2002
  Conference Record of the Thirty-Sixth Asilomar Conference on Signals, Systems
  and Computers, Pacific Grove, CA, USA}}}\ (\bibinfo {year} {IEEE, 2002}),\
  p.\ \bibinfo {pages} {8 vol.1}\BibitemShut {NoStop}%
\bibitem [{\citenamefont {Hocevar}(2004)}]{Hocevar2004}%
  \BibitemOpen
  \bibfield  {author} {\bibinfo {author} {\bibfnamefont {D.~E.}\ \bibnamefont
  {Hocevar}},\ }in\ \href@noop {} {\emph {\bibinfo {booktitle} {2004 IEEE
  Workshop on Signal Processing Systems, Austin, TX, USA}}}\ (\bibinfo {year}
  {IEEE, 2004}),\ p.\ \bibinfo {pages} {107}\BibitemShut {NoStop}%
\bibitem [{\citenamefont {Sharon}\ \emph {et~al.}(2004)\citenamefont {Sharon},
  \citenamefont {Litsyn},\ and\ \citenamefont {Goldberger}}]{Sharon2004An}%
  \BibitemOpen
  \bibfield  {author} {\bibinfo {author} {\bibfnamefont {E.}~\bibnamefont
  {Sharon}}, \bibinfo {author} {\bibfnamefont {S.}~\bibnamefont {Litsyn}}, \
  and\ \bibinfo {author} {\bibfnamefont {J.}~\bibnamefont {Goldberger}},\ }in\
  \href@noop {} {\emph {\bibinfo {booktitle} {2004 IEEE Convention of
  Electrical and Electronics Engineers in Israel, Tel-Aviv, Israel}}}\
  (\bibinfo {year} {IEEE, 2004}),\ p.\ \bibinfo {pages} {223}\BibitemShut
  {NoStop}%
\bibitem [{\citenamefont {Zhang}\ and\ \citenamefont
  {Fossorier}(2004)}]{Zhang2004A}%
  \BibitemOpen
  \bibfield  {author} {\bibinfo {author} {\bibfnamefont {J.}~\bibnamefont
  {Zhang}}\ and\ \bibinfo {author} {\bibfnamefont {M.~P.~C.}\ \bibnamefont
  {Fossorier}},\ }\bibfield  {title} {\bibinfo {title} {A modified weighted
  bit-flipping decoding of low-density parity-check codes},\ }\href@noop {}
  {\bibfield  {journal} {\bibinfo  {journal}
  {\href{https://ieeexplore.ieee.org/document/1278309/}{IEEE Commun. Lett.
  }}}\textbf {\bibinfo {volume}
  {\href{https://ieeexplore.ieee.org/document/1278309/}{8}}}\bibinfo {pages}
  {\href{https://ieeexplore.ieee.org/document/1278309/}{, 165}}\bibinfo {year}
  {\href{https://ieeexplore.ieee.org/document/1278309/}{ (2004).}}}\BibitemShut
  {Stop}%
\bibitem [{\citenamefont {Chang}\ \emph {et~al.}(2008)\citenamefont {Chang},
  \citenamefont {Vila~Casado}, \citenamefont {Chang},\ and\ \citenamefont
  {Wesel}}]{Chang2008Lower}%
  \BibitemOpen
  \bibfield  {author} {\bibinfo {author} {\bibfnamefont {Y.~M.}\ \bibnamefont
  {Chang}}, \bibinfo {author} {\bibfnamefont {A.~I.}\ \bibnamefont
  {Vila~Casado}}, \bibinfo {author} {\bibfnamefont {M.~C.~F.}\ \bibnamefont
  {Chang}}, \ and\ \bibinfo {author} {\bibfnamefont {R.~D.}\ \bibnamefont
  {Wesel}},\ }in\ \href@noop {} {\emph {\bibinfo {booktitle} {2008 IEEE
  International Conference on Communications, Beijing, China}}}\ (\bibinfo
  {year} {IEEE, 2008}),\ p.\ \bibinfo {pages} {1155}\BibitemShut {NoStop}%
\bibitem [{\citenamefont {Park}\ \emph {et~al.}(2008)\citenamefont {Park},
  \citenamefont {Lee},\ and\ \citenamefont {Whang}}]{Park2009Shuffled}%
  \BibitemOpen
  \bibfield  {author} {\bibinfo {author} {\bibfnamefont {S.}~\bibnamefont
  {Park}}, \bibinfo {author} {\bibfnamefont {S.}~\bibnamefont {Lee}}, \ and\
  \bibinfo {author} {\bibfnamefont {K.}~\bibnamefont {Whang}},\ }in\ \href@noop
  {} {\emph {\bibinfo {booktitle} {2008 14th Asia-Pacific Conference on
  Communications, Tokyo, Japan}}}\ (\bibinfo {year} {IEEE, 2008}),\ p.~\bibinfo
  {pages} {1}\BibitemShut {NoStop}%
\bibitem [{\citenamefont {Wu}\ \emph {et~al.}(2010)\citenamefont {Wu},
  \citenamefont {Jiang},\ and\ \citenamefont {Nie}}]{Wu2010Alternate}%
  \BibitemOpen
  \bibfield  {author} {\bibinfo {author} {\bibfnamefont {S.}~\bibnamefont
  {Wu}}, \bibinfo {author} {\bibfnamefont {X.}~\bibnamefont {Jiang}}, \ and\
  \bibinfo {author} {\bibfnamefont {Z.}~\bibnamefont {Nie}},\ }in\ \href@noop
  {} {\emph {\bibinfo {booktitle} {2010 International Conference on
  Communications and Mobile Computing, Shenzhen, China}}}\ (\bibinfo {year}
  {IEEE, 2010}),\ p.\ \bibinfo {pages} {278}\BibitemShut {NoStop}%
\bibitem [{\citenamefont {Aslam}\ \emph {et~al.}(2017)\citenamefont {Aslam},
  \citenamefont {Guan},\ and\ \citenamefont {Cai}}]{Aslam2017Edge}%
  \BibitemOpen
  \bibfield  {author} {\bibinfo {author} {\bibfnamefont {C.~A.}\ \bibnamefont
  {Aslam}}, \bibinfo {author} {\bibfnamefont {Y.~L.}\ \bibnamefont {Guan}}, \
  and\ \bibinfo {author} {\bibfnamefont {K.}~\bibnamefont {Cai}},\ }\bibfield
  {title} {\bibinfo {title} {Edge-based dynamic scheduling for
  belief-propagation decoding of ldpc and rs codes},\ }\href@noop {} {\bibfield
   {journal} {\bibinfo  {journal}
  {\href{https://ieeexplore.ieee.org/abstract/document/7779107/}{IEEE Trans.
  Commun. }}}\textbf {\bibinfo {volume}
  {\href{https://ieeexplore.ieee.org/abstract/document/7779107/}{65}}}\bibinfo
  {pages} {\href{https://ieeexplore.ieee.org/abstract/document/7779107/}{,
  525}}\bibinfo {year}
  {\href{https://ieeexplore.ieee.org/abstract/document/7779107/}{
  (2017).}}}\BibitemShut {Stop}%
\bibitem [{\citenamefont {Kiktenko}\ \emph {et~al.}(2017)\citenamefont
  {Kiktenko}, \citenamefont {Trushechkin}, \citenamefont {Lim}, \citenamefont
  {Kurochkin},\ and\ \citenamefont {Fedorov}}]{kiktenko2017symmetric}%
  \BibitemOpen
  \bibfield  {author} {\bibinfo {author} {\bibfnamefont {E.~O.}\ \bibnamefont
  {Kiktenko}}, \bibinfo {author} {\bibfnamefont {A.~S.}\ \bibnamefont
  {Trushechkin}}, \bibinfo {author} {\bibfnamefont {C.~C.~W.}\ \bibnamefont
  {Lim}}, \bibinfo {author} {\bibfnamefont {Y.~V.}\ \bibnamefont {Kurochkin}},
  \ and\ \bibinfo {author} {\bibfnamefont {A.~K.}\ \bibnamefont {Fedorov}},\
  }\bibfield  {title} {\bibinfo {title} {Symmetric blind information
  reconciliation for quantum key distribution},\ }\href@noop {} {\bibfield
  {journal} {\bibinfo  {journal}
  {\href{https://journals.aps.org/prapplied/abstract/10.1103/PhysRevApplied.8.044017}{Phys.
  Rev. Appl}}}\textbf {\bibinfo {volume}
  {\href{https://journals.aps.org/prapplied/abstract/10.1103/PhysRevApplied.8.044017}{8}}}\bibinfo
  {pages}
  {\href{https://journals.aps.org/prapplied/abstract/10.1103/PhysRevApplied.8.044017}{,
  044017}}\bibinfo {year}
  {\href{https://journals.aps.org/prapplied/abstract/10.1103/PhysRevApplied.8.044017}{
  (2017).}}}\BibitemShut {Stop}%
\bibitem [{\citenamefont {Gallager}(1962)}]{Gallager1962Low}%
  \BibitemOpen
  \bibfield  {author} {\bibinfo {author} {\bibfnamefont {R.~G.}\ \bibnamefont
  {Gallager}},\ }\bibfield  {title} {\bibinfo {title} {Low-density parity-check
  codes},\ }\href@noop {} {\bibfield  {journal} {\bibinfo  {journal}
  {\href{https://ieeexplore.ieee.org/abstract/document/1057683/}{IEEE Trans.
  Inform. Theory. }}}\textbf {\bibinfo {volume}
  {\href{https://ieeexplore.ieee.org/abstract/document/1057683/}{8}}}\bibinfo
  {pages} {\href{https://ieeexplore.ieee.org/abstract/document/1057683/}{,
  3}}\bibinfo {year}
  {\href{https://ieeexplore.ieee.org/abstract/document/1057683/}{
  (1962).}}}\BibitemShut {Stop}%
\bibitem [{\citenamefont {Tanner}(1981)}]{Tanner1981A}%
  \BibitemOpen
  \bibfield  {author} {\bibinfo {author} {\bibfnamefont {R.~M.}\ \bibnamefont
  {Tanner}},\ }\bibfield  {title} {\bibinfo {title} {A recursive approach to
  low complexity codes},\ }\href@noop {} {\bibfield  {journal} {\bibinfo
  {journal} {\href{https://ieeexplore.ieee.org/abstract/document/1056404/}{IEEE
  Trans. Inform. Theory. }}}\textbf {\bibinfo {volume}
  {\href{https://ieeexplore.ieee.org/abstract/document/1056404/}{27}}}\bibinfo
  {pages} {\href{https://ieeexplore.ieee.org/abstract/document/1056404/}{,
  533}}\bibinfo {year}
  {\href{https://ieeexplore.ieee.org/abstract/document/1056404/}{
  (1981).}}}\BibitemShut {Stop}%
\bibitem [{\citenamefont {Mackay}(1999)}]{Mackay1999Good}%
  \BibitemOpen
  \bibfield  {author} {\bibinfo {author} {\bibfnamefont {D.~J.~C.}\
  \bibnamefont {Mackay}},\ }\bibfield  {title} {\bibinfo {title} {Good
  error-correcting codes based on very sparse matrices},\ }\href@noop {}
  {\bibfield  {journal} {\bibinfo  {journal}
  {\href{https://ieeexplore.ieee.org/abstract/document/748992/}{IEEE Trans.
  Inform. Theory. }}}\textbf {\bibinfo {volume}
  {\href{https://ieeexplore.ieee.org/abstract/document/748992/}{47}}}\bibinfo
  {pages} {\href{https://ieeexplore.ieee.org/abstract/document/748992/}{,
  399}}\bibinfo {year}
  {\href{https://ieeexplore.ieee.org/abstract/document/748992/}{
  (1999).}}}\BibitemShut {Stop}%
\bibitem [{\citenamefont {Casado}\ \emph {et~al.}(2007)\citenamefont {Casado},
  \citenamefont {Griot},\ and\ \citenamefont {Wesel}}]{Casado2007Informed}%
  \BibitemOpen
  \bibfield  {author} {\bibinfo {author} {\bibfnamefont {A.~I.~V.}\
  \bibnamefont {Casado}}, \bibinfo {author} {\bibfnamefont {M.}~\bibnamefont
  {Griot}}, \ and\ \bibinfo {author} {\bibfnamefont {R.~D.}\ \bibnamefont
  {Wesel}},\ }in\ \href@noop {} {\emph {\bibinfo {booktitle} {2007 IEEE
  International Conference on Communications, Glasgow, UK}}}\ (\bibinfo {year}
  {IEEE, 2007}),\ p.\ \bibinfo {pages} {107}\BibitemShut {NoStop}%
\bibitem [{\citenamefont {Richardson}\ and\ \citenamefont
  {Urbanke}(2001)}]{richardson2001capacity}%
  \BibitemOpen
  \bibfield  {author} {\bibinfo {author} {\bibfnamefont {T.~J.}\ \bibnamefont
  {Richardson}}\ and\ \bibinfo {author} {\bibfnamefont {R.~L.}\ \bibnamefont
  {Urbanke}},\ }\bibfield  {title} {\bibinfo {title} {The capacity of
  low-density parity-check codes under message-passing decoding},\ }\href@noop
  {} {\bibfield  {journal} {\bibinfo  {journal}
  {\href{https://ieeexplore.ieee.org/abstract/document/910577/}{IEEE Trans.
  Inform. Theory. }}}\textbf {\bibinfo {volume}
  {\href{https://ieeexplore.ieee.org/abstract/document/910577/}{47}}}\bibinfo
  {pages} {\href{https://ieeexplore.ieee.org/abstract/document/910577/}{,
  599}}\bibinfo {year}
  {\href{https://ieeexplore.ieee.org/abstract/document/910577/}{
  (2001).}}}\BibitemShut {Stop}%
\bibitem [{\citenamefont {Richardson}\ \emph {et~al.}(2001)\citenamefont
  {Richardson}, \citenamefont {Shokrollahi},\ and\ \citenamefont
  {Urbanke}}]{richardson2001design}%
  \BibitemOpen
  \bibfield  {author} {\bibinfo {author} {\bibfnamefont {T.~J.}\ \bibnamefont
  {Richardson}}, \bibinfo {author} {\bibfnamefont {M.~A.}\ \bibnamefont
  {Shokrollahi}}, \ and\ \bibinfo {author} {\bibfnamefont {R.~L.}\ \bibnamefont
  {Urbanke}},\ }\bibfield  {title} {\bibinfo {title} {Design of
  capacity-approaching irregular low-density parity-check codes},\ }\href@noop
  {} {\bibfield  {journal} {\bibinfo  {journal}
  {\href{https://ieeexplore.ieee.org/abstract/document/910578/}{IEEE Trans.
  Inform. Theory. }}}\textbf {\bibinfo {volume}
  {\href{https://ieeexplore.ieee.org/abstract/document/910578/}{47}}}\bibinfo
  {pages} {\href{https://ieeexplore.ieee.org/abstract/document/910578/}{,
  619}}\bibinfo {year}
  {\href{https://ieeexplore.ieee.org/abstract/document/910578/}{
  (2001).}}}\BibitemShut {Stop}%
\bibitem [{\citenamefont {Sharon}\ \emph {et~al.}(2007)\citenamefont {Sharon},
  \citenamefont {Litsyn},\ and\ \citenamefont
  {Goldberger}}]{Sharon2007Efficient}%
  \BibitemOpen
  \bibfield  {author} {\bibinfo {author} {\bibfnamefont {E.}~\bibnamefont
  {Sharon}}, \bibinfo {author} {\bibfnamefont {S.}~\bibnamefont {Litsyn}}, \
  and\ \bibinfo {author} {\bibfnamefont {J.}~\bibnamefont {Goldberger}},\
  }\bibfield  {title} {\bibinfo {title} {Efficient serial message-passing
  schedules for ldpc decoding},\ }\href@noop {} {\bibfield  {journal} {\bibinfo
   {journal}
  {\href{https://ieeexplore.ieee.org/abstract/document/4373433/}{IEEE Trans.
  Inform. }}}\textbf {\bibinfo {volume}
  {\href{https://ieeexplore.ieee.org/abstract/document/4373433/}{53}}}\bibinfo
  {pages} {\href{https://ieeexplore.ieee.org/abstract/document/4373433/}{,
  4076}}\bibinfo {year}
  {\href{https://ieeexplore.ieee.org/abstract/document/4373433/}{
  (2007).}}}\BibitemShut {Stop}%
\bibitem [{\citenamefont {Casado}\ \emph {et~al.}(2010)\citenamefont {Casado},
  \citenamefont {Griot},\ and\ \citenamefont {Wesel}}]{Casado2010LDPC}%
  \BibitemOpen
  \bibfield  {author} {\bibinfo {author} {\bibfnamefont {A.~I.~V.}\
  \bibnamefont {Casado}}, \bibinfo {author} {\bibfnamefont {M.}~\bibnamefont
  {Griot}}, \ and\ \bibinfo {author} {\bibfnamefont {R.~D.}\ \bibnamefont
  {Wesel}},\ }\bibfield  {title} {\bibinfo {title} {Ldpc decoders with informed
  dynamic scheduling},\ }\href@noop {} {\bibfield  {journal} {\bibinfo
  {journal} {\href{https://ieeexplore.ieee.org/abstract/document/5610969/}{IEEE
  Trans. Commun. }}}\textbf {\bibinfo {volume}
  {\href{https://ieeexplore.ieee.org/abstract/document/5610969/}{58}}}\bibinfo
  {pages} {\href{https://ieeexplore.ieee.org/abstract/document/5610969/}{,
  3470}}\bibinfo {year}
  {\href{https://ieeexplore.ieee.org/abstract/document/5610969/}{
  (2010)}}}\BibitemShut {NoStop}%
\bibitem [{\citenamefont {Yazdani}\ \emph {et~al.}(2004)\citenamefont
  {Yazdani}, \citenamefont {Hemati},\ and\ \citenamefont
  {Banihashemi}}]{Yazdani2004Improving}%
  \BibitemOpen
  \bibfield  {author} {\bibinfo {author} {\bibfnamefont {M.~R.}\ \bibnamefont
  {Yazdani}}, \bibinfo {author} {\bibfnamefont {S.}~\bibnamefont {Hemati}}, \
  and\ \bibinfo {author} {\bibfnamefont {A.~H.}\ \bibnamefont {Banihashemi}},\
  }\bibfield  {title} {\bibinfo {title} {Improving belief propagation on graphs
  with cycles},\ }\href@noop {} {\bibfield  {journal} {\bibinfo  {journal}
  {\href{https://ieeexplore.ieee.org/abstract/document/1261926/}{IEEE Commun.
  Lett. }}}\textbf {\bibinfo {volume}
  {\href{https://ieeexplore.ieee.org/abstract/document/1261926/}{8}}}\bibinfo
  {pages} {\href{https://ieeexplore.ieee.org/abstract/document/1261926/}{,
  57}}\bibinfo {year}
  {\href{https://ieeexplore.ieee.org/abstract/document/1261926/}{
  (2004).}}}\BibitemShut {Stop}%
\bibitem [{\citenamefont {Zhang}\ and\ \citenamefont
  {Pfister}(2012)}]{Zhang2012Verification}%
  \BibitemOpen
  \bibfield  {author} {\bibinfo {author} {\bibfnamefont {F.}~\bibnamefont
  {Zhang}}\ and\ \bibinfo {author} {\bibfnamefont {H.~D.}\ \bibnamefont
  {Pfister}},\ }\bibfield  {title} {\bibinfo {title} {Verification decoding of
  high-rate ldpc codes with applications in compressed sensing},\ }\href@noop
  {} {\bibfield  {journal} {\bibinfo  {journal}
  {\href{https://ieeexplore.ieee.org/abstract/document/6205393/}{IEEE Trans.
  Inform. }}}\textbf {\bibinfo {volume}
  {\href{https://ieeexplore.ieee.org/abstract/document/6205393/}{58}}}\bibinfo
  {pages} {\href{https://ieeexplore.ieee.org/abstract/document/6205393/}{,
  5042}}\bibinfo {year}
  {\href{https://ieeexplore.ieee.org/abstract/document/6205393/}{
  (2012).}}}\BibitemShut {Stop}%
\bibitem [{\citenamefont {Djordjevic}\ \emph {et~al.}(2012)\citenamefont
  {Djordjevic}, \citenamefont {Arabaci},\ and\ \citenamefont
  {Zhang}}]{Djordjevic2012Evaluation}%
  \BibitemOpen
  \bibfield  {author} {\bibinfo {author} {\bibfnamefont {I.~B.}\ \bibnamefont
  {Djordjevic}}, \bibinfo {author} {\bibfnamefont {M.}~\bibnamefont {Arabaci}},
  \ and\ \bibinfo {author} {\bibfnamefont {Y.}~\bibnamefont {Zhang}},\
  }\bibfield  {title} {\bibinfo {title} {Evaluation of four-dimensional
  nonbinary ldpc-coded modulation for next-generation long-haul optical
  transport networks},\ }\href@noop {} {\bibfield  {journal} {\bibinfo
  {journal} {\href{https://europepmc.org/abstract/med/22513641}{Opt. Express
  }}}\textbf {\bibinfo {volume}
  {\href{https://europepmc.org/abstract/med/22513641}{20}}}\bibinfo {pages}
  {\href{https://europepmc.org/abstract/med/22513641}{, 9296}}\bibinfo {year}
  {\href{https://europepmc.org/abstract/med/22513641}{ (2012).}}}\BibitemShut
  {Stop}%
\bibitem [{\citenamefont {Luby}\ \emph {et~al.}(2001)\citenamefont {Luby},
  \citenamefont {Mitzenmacher}, \citenamefont {Shokrollahi},\ and\
  \citenamefont {Spielman}}]{Luby2001Improved}%
  \BibitemOpen
  \bibfield  {author} {\bibinfo {author} {\bibfnamefont {M.~G.}\ \bibnamefont
  {Luby}}, \bibinfo {author} {\bibfnamefont {M.}~\bibnamefont {Mitzenmacher}},
  \bibinfo {author} {\bibfnamefont {M.~A.}\ \bibnamefont {Shokrollahi}}, \ and\
  \bibinfo {author} {\bibfnamefont {D.~A.}\ \bibnamefont {Spielman}},\
  }\bibfield  {title} {\bibinfo {title} {Improved low-density parity-check
  codes using irregular graphs},\ }\href@noop {} {\bibfield  {journal}
  {\bibinfo  {journal}
  {\href{https://ieeexplore.ieee.org/iel5/18/19638/910576/910576.plain.html}{IEEE
  Trans. Inform. Theory. }}}\textbf {\bibinfo {volume}
  {\href{https://ieeexplore.ieee.org/iel5/18/19638/910576/910576.plain.html}{47}}}\bibinfo
  {pages}
  {\href{https://ieeexplore.ieee.org/iel5/18/19638/910576/910576.plain.html}{,
  585}}\bibinfo {year}
  {\href{https://ieeexplore.ieee.org/iel5/18/19638/910576/910576.plain.html}{
  (2001).}}}\BibitemShut {Stop}%
\end{thebibliography}%

\appendix
\section{Appendix A\\ Security Analysis}

The security of single-matrix reconciliation is guaranteed by the following theorems.

\textbf{Theorem 1}: Let $x$ and $z$ be Alice's sifted key and syndrome, respectively. $H_{m\times n}$ is the matrix used in reconciliation. Once Eve gets $z$, she can extract at most $m$ bits of information about $x$, i.e.,
\begin{equation}
I(x;z)\leq m.
\label{equ:prove-1}
\end{equation}

\noindent \textbf{Proof of Theorem}: The amount of information that Eve can obtain from $z$ about $x$ is
\begin{equation}
I(x;z)=H(z)-H(z|x).
\label{equ:prove-2}
\end{equation}

\noindent Assuming that Eve knows $H_{m\times n}$, she would obtain $z$ if she knows $x$, i.e.,
\begin{equation}
H(z|x)=0.
\label{equ:prove-3}
\end{equation}

\noindent When a random variables are in the equal probability distribution, the discrete entropy can reach the maximum value, so
\begin{equation}
\begin{split}
I(x;z)&=H(z)\leq -\sum_1^{2^{m}}(\frac{1}{2^{m}}\log{\frac{1}{2^{m}}})\\
&=\log{2^{m}}=m.
\end{split}
\label{equ:prove-4}
\end{equation}

\textbf{Theorem 2}: If the random variable $X$ of Alice's sifted key $x$ obeys uniform distribution, i.e.,
\begin{equation}
P(X=x)=\frac{1}{2^{n}},
\label{equ:prove-5}
\end{equation}

\noindent then there are at least $t$ bits of information about $x$ unknown to Eve, even though she has obtained $z$, i.e.,
\begin{equation}
H(x|z)\ge t.
\label{equ:prove-6}
\end{equation}
\noindent where $t = n - m $.

\noindent \textbf{Proof of Theorem}: The random variable $X$ of Alice's sifted key obeys uniform distribution, so
\begin{equation}
H(x)=-\sum_1^{2^{n}}(\frac{1}{2^{n}}\log{\frac{1}{2^{n}}})=\log{2^{n}}=n.
\label{equ:prove-7}
\end{equation}

\noindent From equations (\ref{equ:prove-1}), (\ref{equ:prove-2}), and (\ref{equ:prove-3}), we derive
\begin{equation}
\begin{split}
H(x|z)&=H(x)-H(z)\\
&=H(x)-I(x;z)\ge n-m=t.
\label{equ:prove-8}
\end{split}
\end{equation}

According to \textbf{Theorem 1} and \textbf{Theorem 2}, Eve can get at most $m$ bits information. Thus, if the $m$ bits  is discarded
during privacy amplification, the security of the key can be guaranteed.

Generally, Alice and Bob can use the following method to abandon the $m$ bits information leakage.
If the matrix $H_{m\times n}$ has the following structure,
\begin{equation}
H_{m\times n}=(H_{m\times t}^{'},E_m),
\label{equ:prove-9}
\end{equation}

\noindent where $H_{m\times t}^{'}$ is a matrix which has $m$ rows and $t$ columns, $E_m$ is an m-order identity matrix, then $H_{m\times n}$ is called a system code.
In other words, $m$ vectors of $E_m$ are linearly independent in $H_{m\times n}$. Under this circumstance, Alice can calculate and send the syndrome by
\begin{equation}
\begin{split}
z&=(H_{m\times t}^{'},E_m)\cdot x\\
&=H_{m\times t}^{'}\cdot\begin{bmatrix} x_1 \\ x_2 \\ \vdots \\ x_t \end{bmatrix}\oplus\begin{bmatrix} x_{t+1} \\ x_{t+2} \\ \vdots \\ x_n \end{bmatrix}=\begin{bmatrix} z_1 \\ z_2 \\ \vdots \\ z_m \end{bmatrix}.
\label{equ:prove-10}
\end{split}
\end{equation}

\noindent From \textbf{Theorem 1}, we know that Eve can obtain at most $m$ bits of information about $x$.
Assume these $m$ bits of information is $m$ bits of $x$.
And for Eve, it is in her best interests if the $m$ bits of $x$ are $[x_{t+1}, \cdots, x_n]^{T}$.
Then Eve has to solve a underdetermined system of equation, which has no unique solution.
Moreover, after Alice and Bob discard the $m$ bits key $[x_{t+1}, \cdots, x_n]^{T}$, Eve cannot even form the system of equation and get any information about $[x_1, \cdots, x_t]^{T}$, even if she knows $H_{m\times t}^{'}$ and $z$.

If the matrix $H_{m\times n}$ is a non-system code, a system code can be formed by a series of elementary row transformations and column exchanges based on
\begin{equation}
H_{m\times n}=A\cdot(H_{m\times t}^{'},E_m)\cdot B,
\label{equ:prove-11}
\end{equation}

\noindent where A is a m-order invertible square matrix representing a whole train of primary row transformation. B is a n-order square matrix representing a series of column exchanges. Denote ${z^{'}}^{T}={(A^{-1}\cdot z)}^{T}=[z_1^{'},   \cdots, z_m^{'}]$ and ${x^{'}}^{T}={(B\cdot x)}^{T}=[x_1^{'}, \cdots, x_n^{'}]$, then
\begin{equation}
\begin{split}
z^{'}&=(H_{m\times t}^{'},E_m)\cdot x^{'}\\
&=H_{m\times t}^{'}\cdot\begin{bmatrix} x_1^{'} \\ x_2^{'} \\ \vdots \\ x_t^{'} \end{bmatrix}\oplus\begin{bmatrix} x_{t+1}^{'} \\ x_{t+2}^{'} \\ \vdots \\ x_n^{'} \end{bmatrix}=\begin{bmatrix} z_1^{'} \\ z_2^{'} \\ \vdots \\ z_m^{'} \end{bmatrix}.
\label{equ:prove-12}
\end{split}
\end{equation}

\noindent Similarly, after Alice and Bob abandon the $m$ bits key $[x_{t+1}^{'}, \cdots, x_n^{'}]^{T}$, even if Eve knows $H_{m\times t}^{'}$ and $z^{'}$, she will not be able to get any information about $[x_1^{'}, \cdots, x_t^{'}]^{T}$.

From the above analysis, we can see that if we first select $m$ linearly independent columns in $H_{m\times n}$, then discard the corresponding bits of these columns, the $m$ bits information leakage can be removed, thus ensuring the security of the key. Therefore, we design a multiple matrices construction method as shown in the Appendix B. And all matrices used in the simulation are prepared according to the the method.

Through the above method, we can construct a series of matrices $(H_1, \cdots, H_u)$ of the same size. Let $H_i$ and $H_j$ denote any two matrices from $(H_1, \cdots, H_u)$. They can be represented as follows:
\begin{equation}
H_i=A_i\cdot (H_i^{'},E_m)\cdot B_i,
\label{equ:prove-13}
\end{equation}
\begin{equation}
H_j=A_j\cdot (H_j^{'},E_m)\cdot B_j,
\label{equ:prove-14}
\end{equation}

\noindent Their syndromes $z^{'i}$ and $z^{'j}$ can be represented as:

\begin{equation}
\begin{split}
z^{'i}&=(H_{i}^{'},E_m)\cdot x^{'i}\\
&=H_{i}^{'}\cdot\begin{bmatrix} x_{1}^{'i} \\ x_{2}^{'i} \\ \vdots \\ x_{t}^{'i} \end{bmatrix}\oplus\begin{bmatrix} x_{t+1}^{'i} \\ x_{t+2}^{'i} \\ \vdots \\ x_{n}^{'i} \end{bmatrix}=\begin{bmatrix} z_{1}^{'i} \\ z_{2}^{'i} \\ \vdots \\ z_{m}^{'i} \end{bmatrix},
\label{equ:prove-15}
\end{split}
\end{equation}
\begin{equation}
\begin{split}
z^{'j}&=(H_{j}^{'},E_m)\cdot x^{'j}\\
&=H_{j}^{'}\cdot\begin{bmatrix} x_{1}^{'j} \\ x_{2}^{'j} \\ \vdots \\ x_{t}^{'j} \end{bmatrix}\oplus\begin{bmatrix} x_{t+1}^{'j} \\ x_{t+2}^{'j} \\ \vdots \\ x_{n}^{'j} \end{bmatrix}=\begin{bmatrix} z_{1}^{'j} \\ z_{2}^{'j} \\ \vdots \\ z_{m}^{'j} \end{bmatrix},
\label{equ:prove-16}
\end{split}
\end{equation}

\noindent  More precisely, ${z^{'i}}^{T}={(A_i^{-1}\cdot z^{i})}^{T}=[z_{1}^{'i},\cdots,z_{m}^{'i}]$, ${x^{'i}}^{T}={(B_i\cdot x)}^{T}=[x_{1}^{'i},\cdots,x_{n}^{'i}]$, ${z^{'j}}^{T}={(A_j^{-1}\cdot z^{j})}^{T}=[z_{1}^{'j},\cdots,z_{m}^{'j}]$, and ${x^{'j}}^{T}={(B_j\cdot x)}^{T}=[x_{1}^{'j},\cdots,x_{n}^{'j}]$. From the above matrices construction method, we can see that $[x_{t+1}^{'i},\cdots,x_{n}^{'i}]$ and $[x_{t+1}^{'j},\cdots,x_{n}^{'j}]$ are not equal, but their corresponding variable nodes sets are the same.
Similarly, assume Eve knows $[x_{t+1}^{'i},\cdots,x_{n}^{'i}]$ and $[x_{t+1}^{'j},\cdots,x_{n}^{'j}]$, then she has to solve the system of equation.
Because $H_i$ and $H_j$ are construct with the method in the Appendix B, the two sets of underdetermined systems of equation in equations (\ref{equ:prove-15}) and (\ref{equ:prove-16}) are the same.
In other word, it is impossible to form a determined or overdetermined system of equation.
After Alice and Bob discard those $m$ bits, even if Eve knows $H_i$, $H_j$, $z^{i}$, and $z^{j}$, she cannot obtain any information about $[x_{1}^{'i},\cdots,x_{t}^{'i}]$ and $[x_{1}^{'j},\cdots,x_{t}^{'j}]$. In fact, any two matrices constructed by this method will not reveal extra information during reconciliation.
Accordingly, in the case of reconciliation with more than two matrices, because the discarded $m$ bits information is corresponding to the same $m$ linearly independent columns, multiple syndromes transmitted through the classical channel do not reveal extra information, i.e.
\begin{equation}
H(z^{i}|z^{i-1},\cdots,z^{1})=0,
\forall z^{i}\in \{z^{2},\cdots,z^{u}\},
\label{equ:prove-17}
\end{equation}

\noindent then we get
\begin{equation}
\begin{aligned}
I(x;Z)&=I(x;z^{1})+I(x;z^{2}|z^{1})+I(x;z^{3}|z^{2} z^{1})+ \\
&\cdots+I(x;z^{u}|z^{u-1} \cdots\ z^{1}) \\
&=H(z^{1})-H(z^{1}|x)+H(z^{2}|z^{1})-H(z^{2}|x\ z^{1})+ \\
&\cdots+H(z^{u}|z^{u-1}\cdots z^{1})-H(z^{u}|x\ z^{u-1}\cdots z^{1}) \\
&=H(z^{1})+H(z^{2}|z^{1})+\cdots +H(z^{u}|z^{u-1}\cdots z^{1}) \\
&=H(z^{1})
\end{aligned}
\label{equ:prove-18}
\end{equation}

\noindent where $Z=\{z^{1},z^{2},\cdots,z^{u}\}$. Therefore, if Alice and Bob use our method to construct matrices, they can guarantee the security of the key, i.e., guarantee the security of the multi-matrix post-processing.

\section{Appendix B\\ Multiple Matrices Construction Method}
\begin{enumerate}
\item The first LDPC matrix called $H_1$ is constructed;
\item By a series of elementary row transformation and column exchanges, $H_1$ is transformed into a system code, such that $m$ linearly independent columns can be determined.
These columns correspond to variable nodes $[v_{(1)}, \cdots, v_{(m)}]$ in $H_1$. Let the remaining variable nodes be $[v_{[1]}, \cdots, v_{[t]}]$;
\item The rest $u-1$ parity check matrices $(H_2, \cdots ,H_u)$ can be constructed based on $H_1$: First, rearrange the columns of the variable nodes $[v_{(1)}, \cdots, v_{(m)}]$ in $H_1$. Then rearrange the columns of the variable nodes $[v_{[1]}, \cdots, v_{[t
    ]}]$ in $H_1$. It's clear that the set of the positions of linearly independent columns, in this way, is identical to each other for all of the $u$ matrices.
\end{enumerate}

\section{Appendix C\\ Pseudocode of MBP, MSBP and MLBP}

\textbf{MBP algorithm}
\begin{algorithmic}[1]
 	\State $Initialize\ L_{v_i\to c_j}^{k}=L_{P_i}^{k}$
 	\For{$every\ parity\textrm{-}check\ matrix\ H_k$}
        \For{$j=1$ to $m$}
            \For{$every\ v_i^{k}\in neighborhood\ of\ c_j^{k}$}
                \State $Generate\ and\ propagate\ L_{c_j\to v_i}^{k}$
            \EndFor
        \EndFor
        \For{$i=1$ to $n$}
            \For{$every\ c_j^{k}\in neighborhood\ of\ v_i^{k}$}
                \State $Generate\ and\ propagate\ L_{v_i\to c_j}^{k}$
            \EndFor
        \EndFor
	\EndFor
    \State $Make\ decoding\ decisions$
    \If{$ stopping\ rule\ is\ not\ satisfied$}
        \State $Go\ back\ to\ line\ 2$
    \EndIf
\end{algorithmic}

~\\
\textbf{MSBP algorithm}
\begin{algorithmic}[1]                             
 	\State $Initialize\ L_{v_i\to c_j}^{k}=L_{P_i}^{k}$
 	\For{$every\ parity\textrm{-}check\ matrix\ H_k$}
        \For{$i=1$ to $n$}
            \For{$every\ c_j^{k}\in neighborhood\ of\ v_i^{k}$}
                \State $Generate\ and\ propagate\ L_{c_j\to v_i}^{k}$
            \EndFor
            \For{$every\ c_j^{k}\in neighborhood\ of\ v_i^{k}$}
                \State $Generate\ and\ propagate\ L_{v_i\to c_j}^{k}$
            \EndFor
        \EndFor
	\EndFor
    \State $Make\ decoding\ decisions$
    \If {$stopping\ rule\ is\ not\ satisfied$}
        \State $Go\ back\ to\ line\ 2$
    \EndIf
\end{algorithmic}

~\\
\textbf{MLBP algorithm}
\begin{algorithmic}[1]                            
	\State $Initialize\ L_{v_i\to c_j}^{k}=L_{P_i}^{k}$
	\For{$every\ parity\textrm{-}check\ matrix\ H_k$}
        \For{$j=1$ to $m$}
            \For{$every\ v_i^{k}\in neighborhood\ of\ c_j^{k}$}
                \State $Generate\ and\ propagate\ L_{c_j\to v_i}^{k}$
                \For{$every\ c_l^{k}\in neighborhood of\ v_i^{k}\ except\ c_j^{k}$}
                    \State $Generate\ and\ propagate\ L_{v_i\to c_l}^{k}$
                \EndFor
            \EndFor
        \EndFor
	\EndFor
    \State  $Make\ decoding\ decisions$
    \If{$sto pping\ rule\ is\ not\ satisfied$}
        \State $Go\ back\ to\ line\ 2$
    \EndIf
\end{algorithmic}

\end{document}